\documentclass[aps,prd,preprint,nofootinbib]{revtex4-1}

\usepackage{graphicx}
\usepackage{epstopdf}
\usepackage{subfigure}
\usepackage{dcolumn}
\usepackage{amsmath}

 \usepackage[font=footnotesize]{caption}

\begin{document}
\title{Suggestion of Coherent Radio Reflections from an Electron-Beam Induced Particle Cascade}

\author{S Prohira$^{1,2}$}
\author{K.~D.~de~Vries$^3$}
\author{D.~Besson$^{4,9}$}
\author{A.~Connolly$^{1,2}$}
\author{C.~Hast$^{5}$}
\author{U.~Latif$^{4}$}
\author{T.~Meures$^{6}$}
\author{J.P.~Ralston$^{4}$}
\author{Z.~Riesen$^{7}$}
\author{D.~Saltzberg$^{8}$}
\author{J.~Torres$^1$}
\author{S.~Wissel$^7$}
\author{X.~Zuo$^8$}

\affiliation{$^1$The Ohio State University, Columbus, OH, USA}
\affiliation{$^2$Center for Cosmology and AstroParticle Physics, The Ohio State University, Columbus, OH, USA}
\affiliation{$^3$Vrije University Brussels, Brussels, Belgium}
\affiliation{$^4$University of Kansas, Lawrence, KS, USA}
\affiliation{$^5$SLAC National Accelerator Laboratory, Menlo Park, CA, USA}
\affiliation{$^6$University of Wisconsin-Madison, Madison, WI, USA}

\affiliation{$^7$California Polytechnic State University, San Luis Obispo, CA, USA}
\affiliation{$^8$University of California-Los Angeles, Los Angeles, CA, USA}
\affiliation{$^9$National Research Nuclear University, Moscow Engineering Physics Institute, Moscow, Russia}



\date{\today}

\begin{abstract}
Testbeam experiment 576 (T576) at the SLAC National Accelerator Laboratory sought to make the first measurement of coherent radio reflections from the ionization produced in the wake of a high-energy particle shower. The $>$10~GeV electron beam at SLAC End Station A was directed into a large high-density polyethylene target to produce a shower analogous to that produced by an EeV neutrino interaction in ice. Continuous wave (CW) radio was transmitted into the target, and receiving antennas monitored for reflection of the transmitted signal from the ionization left in the wake of the shower. We detail the experiment and report on preliminary hints of a signal consistent with a radio reflection at a significance of $2.36\sigma$. We recommend another test-beam measurement in order to verify the signal.
\end{abstract}

\maketitle
\section{Introduction}

A particle shower in a medium produces high energy particles that traverse that medium, ejecting ionization electrons from atoms in the bulk as the shower evolves. For high enough incident energies, this ionization may become dense enough to reflect at radio wavelengths, approximating a short-lived, cylindrical conductor. The Telescope Array RAdar (TARA)~\cite{tara} project was the first dedicated experiment to attempt detection of the extensive air shower (EAS)\cite{eas} from a cosmic ray interaction in the atmosphere using the radar method. TARA reported no signal~\cite{tara_limit}, but placed a strong experimental limit on the extant model of in-air radar reflections~\cite{gorham}. Several experiments have sought to detect radar reflections from ionization deposits in a laboratory setting~\cite{chiba,krijn_els_1,krijn_els}. The Chiba~\cite{chiba} group reported positive results for reflections from ionization deposits in dense material, albeit not from particle shower-induced ionization. The T576 experiment at SLAC was designed to make the first direct measurement of radar reflection from the ionization produced by a particle shower.  

The $\cal{O}$(1-10 GeV) electron beam at SLAC has a nominal bunch number of $10^9$ electrons. Directing this beam into a target of high-density polyethylene (HDPE) produces a shower equivalent to that produced by a 1~EeV primary neutrino, which can be interrogated with radio in an effort to quantify the ionization parameters of a true neutrino-induced cascade. To that end, Testbeam experiment 576, or T576, ran in May 2018. 

\section{Experimental Setup}
The End Station Test Beam facility provides users a $\cal{O}$(1~Hz) bunch of high energy electrons switched from the main linear accelerator (linac) over into End Station A (ESA). ESA is a `parasitic' user facility at SLAC; i.e., the parameters of the electron bunch (energy, beam current) are selected by the main linac user, rather than the End Station user. For our purposes this was actually advantageous, as a scan of energies and currents allowed investigation of how a putative signal depends on those parameters. For T576 the beam current was typically $\sim$250~pC, corresponding to roughly $10^9$ electrons per bunch. The run-time variation of the beam current is shown in Figure~\ref{ict_dist}. The primary electron energy varied from 10--14.4~GeV throughout the experiment, with most of the data accumulated at 14.4~GeV. At the point where the beam exits the beam pipe at the end of the ESA, the bunch is highly collimated, occupying less than a cubic centimeter in volume.

\begin{figure}
\begin{centering}
\includegraphics[width=.7\textwidth]{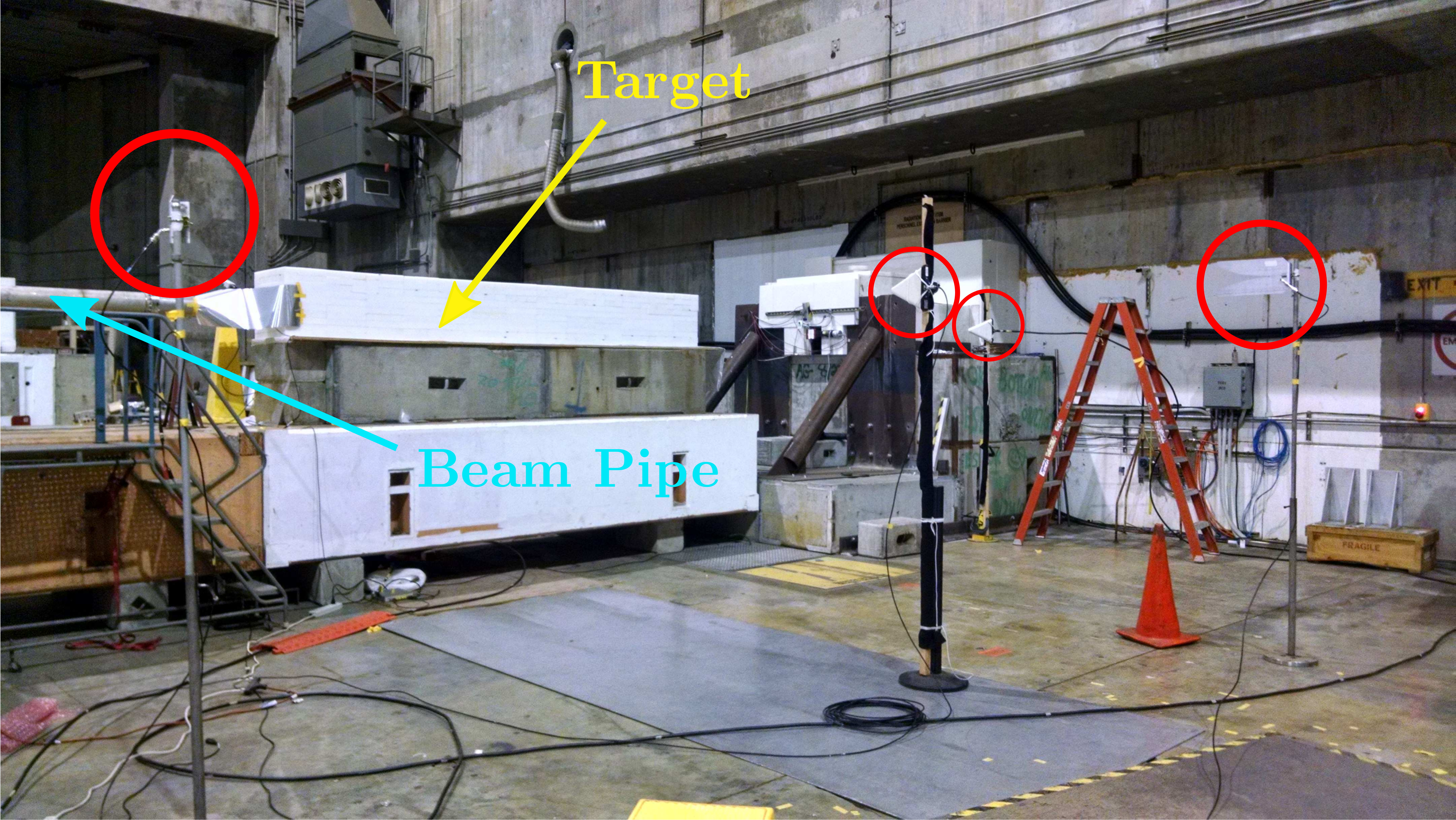}
\par\end{centering}
\caption{The T576 experimental setup. The large white rectangular polyhedron at the center is the HDPE target. The beam enters from the left, with the entry point shielded by aluminum sheeting in an effort to mitigate transition radiation (TR). The circles (red online) indicate the receiver/transmitter antennas. Second from left is the transmitter, the others are receivers.}
\label{t576setup}
\end{figure} 

Figure~\ref{t576setup} shows the target assembled on-site at End Station A at SLAC. The HDPE target was initially constructed for the T510~\cite{t510} experiment, for which it was used to study the geomagnetic emission from a particle shower created within the plastic target. For T576, the HDPE target was aligned with the beam by placing it on top of large concrete blocks. Transmitting and receiving antennas were positioned around the target in various configurations throughout the experiment, as described in detail below. Two different types of antennas were used: an LPDA having voltage standing-wave ratio (VSWR) less than 3.0 over a 1-18~GHz bandwidth, and a Vivaldi antenna with a 0.6-6~GHz bandwidth. The transmitter and receiver amplification was varied throughout the experiment as well, in order to quantify and mitigate backgrounds and also investigate the scaling properties of observed signals. 

\begin{figure}
\begin{centering}
\includegraphics[width=.7\textwidth]{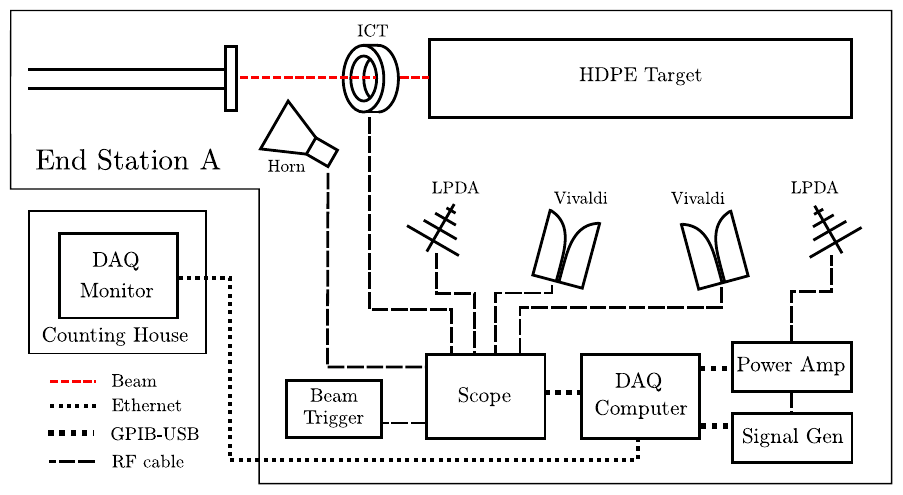}
\par\end{centering}
\caption{The T576 signal chain. The DAQ system and transmitter resided in the End Station A, and were remotely monitored via an Ethernet link from the `counting house', a remotely accessible location for users while the beam is on. The various components shown are described in the text. }
\label{signal_chain}
\end{figure} 

A typical signal chain and the DAQ configuration are presented in Figure~\ref{signal_chain}. In the later analysis section, more detail will be given on the amplification and filtration choices made for various `runs' during the experiment. The DAQ was a Tektronix TDS-694C 4-channel, 10~GS/s digital oscilloscope, connected to a laptop via a GPIB-USB adapter. This laptop was remotely accessible via a network link from the `counting house', which allowed for control of all scope parameters and real-time readout of the data. The transmitter, a Rhode and Schwartz SMHU signal generator, was also controlled remotely via the same computer with another GPIB-USB adapter, allowing real-time frequency and output level tuning. The final piece of equipment (also controlled via GPIB cable) was an Instruments for Industry SMCC100 power amplifier for the transmitter, permitting output level variation, as well as automatic levelling control and queries for forward and reflected power. As no personnel are allowed inside of the End Station during operation, having such a high degree of remote control over the parameters of the experiment was critical for minimizing down-time for hardware adjustments, thereby allowing accumulation of as much data as possible. An integrating current toroid (ICT) was used to monitor the beam current for every event, and occupied the 4th channel of the scope for the duration of the experiment. 

The scope was triggered by either a) a logic pulse from the accelerator itself or b) a sharp transition radiation signal from an s-band horn (indicated by `horn' in Figure~\ref{signal_chain}), depending on the run. 
The TR horn signal was very sharp and consistent, but most of our data was taken using the beam logic pulse as a trigger since it could be modified remotely to allow precise time shifts of trigger point relative to the true arrival of the beam.
Later runs substituted a third receiver antenna for the s-band horn, to better characterize the expected reflection signal as a function of angle.  

Part-way through the run, the reported power amplifier output level began to drift by approximately 20\% compared to the actual output power (determined by observing signal strength in the scope). In what follows, we therefore assume a 20\% systematic error on the transmitter output power.

\section{Specific Measurements and Goals}
For T576 and the radar problem, there are model-dependent and model-independent measurements that can be made. Model-dependent measurements in this case correspond to those observables which are dependent on multiple parameters, e.g., the plasma lifetime and the microscopic scattering physics. The spectral content, temporal duration and angular distribution of signal are important observables that will ideally either falsify or confirm different models. Simple measurements of coherence~\cite{bean_ralston}, that is, whether the received power in the signal region scales with distance as $R^{-4}$, and linearly with transmitter power, are considered model-independent measurements.\footnote{The $R^{-4}$ scaling derives directly from the radar equation, which more specifically prescribes a scaling $\propto R_1^2R_2^2$, where $R_1$ is the distance from transmitter to reflector and $R_2$ is the distance from reflector to receiver. For a fixed baseline, the received power should scale linearly with transmitter power.} 
For T576, we attempted as many different combinations of measurements as possible to test our results in both model-dependent and independent manners.

\begin{figure}
\begin{centering}
\includegraphics[width=.5\textwidth]{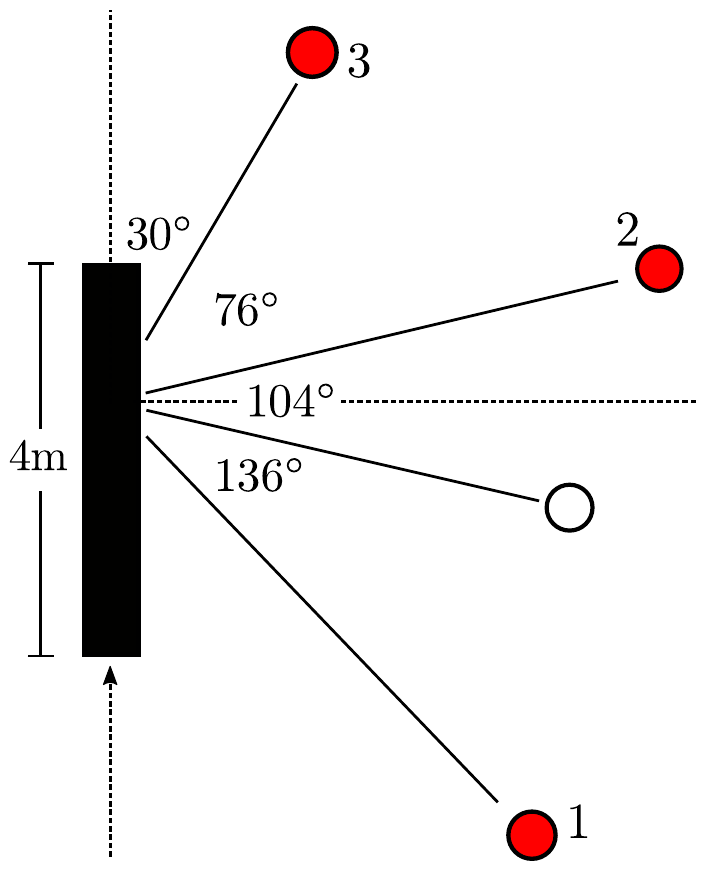}
\par\end{centering}
\caption{The setup for run 11 of T576, viewed from above, and drawn to scale. Closed circles are receivers, labeled by their DAQ channel number, open circle is the transmitter. One receiver is at the specular reflection point relative to calculated shower maximum; the others are separated from the specular angles by 30--40 degrees.}
\label{setup_run_11}
\end{figure}

Figure~\ref{setup_run_11} shows the configuration for one run. In this run, all three receivers are positioned on one side of the target. One receiver was positioned at the specular reflection point for shower maximum (calculated to be roughly 3~m longitudinally into the target), with the other two set off at either side. This provides a model-dependent measurement. A large stationary conductor in the target region was positioned so as to reflect at the specular point; reflected amplitudes should decrease sharply as the receivers move away from this point, assuming a sufficiently long plasma lifetime. If, however, the plasma lifetime is short, then the majority of scattering may be in the single-particle regime, which is far more isotropic. Observation of reflection in the non-specular receivers would therefore point to a short plasma lifetime. 

The overall goal of T576 was to measure an unambiguous radar reflection from a particle shower. The next sections will discuss the challenges to making such a measurement and the analysis of the data. As will be discussed, the extremely high backgrounds made many of these goals difficult to attain. However, after applying a particularly sensitive method for small-signal detection in large backgrounds, we present strong suggestions of a signal which warrant further investigation.

\section{Backgrounds}
There were several backgrounds at ESA during T576 data-taking. The typical radio-frequency (RF) backgrounds, anthropogenic and generic low-level electromagnetic interference (EMI), were relatively low within the thick concrete bunker-style building of the ESA. Occasional bursts of communications radio were observed, but well-below trigger threshold. The two most pernicious backgrounds were observed to be room reflections and a very strong RF signal from the beam/target interaction itself. 

\subsection{Spurious Reflections}
\begin{figure}
\begin{centering}
\includegraphics[width=.6\textwidth]{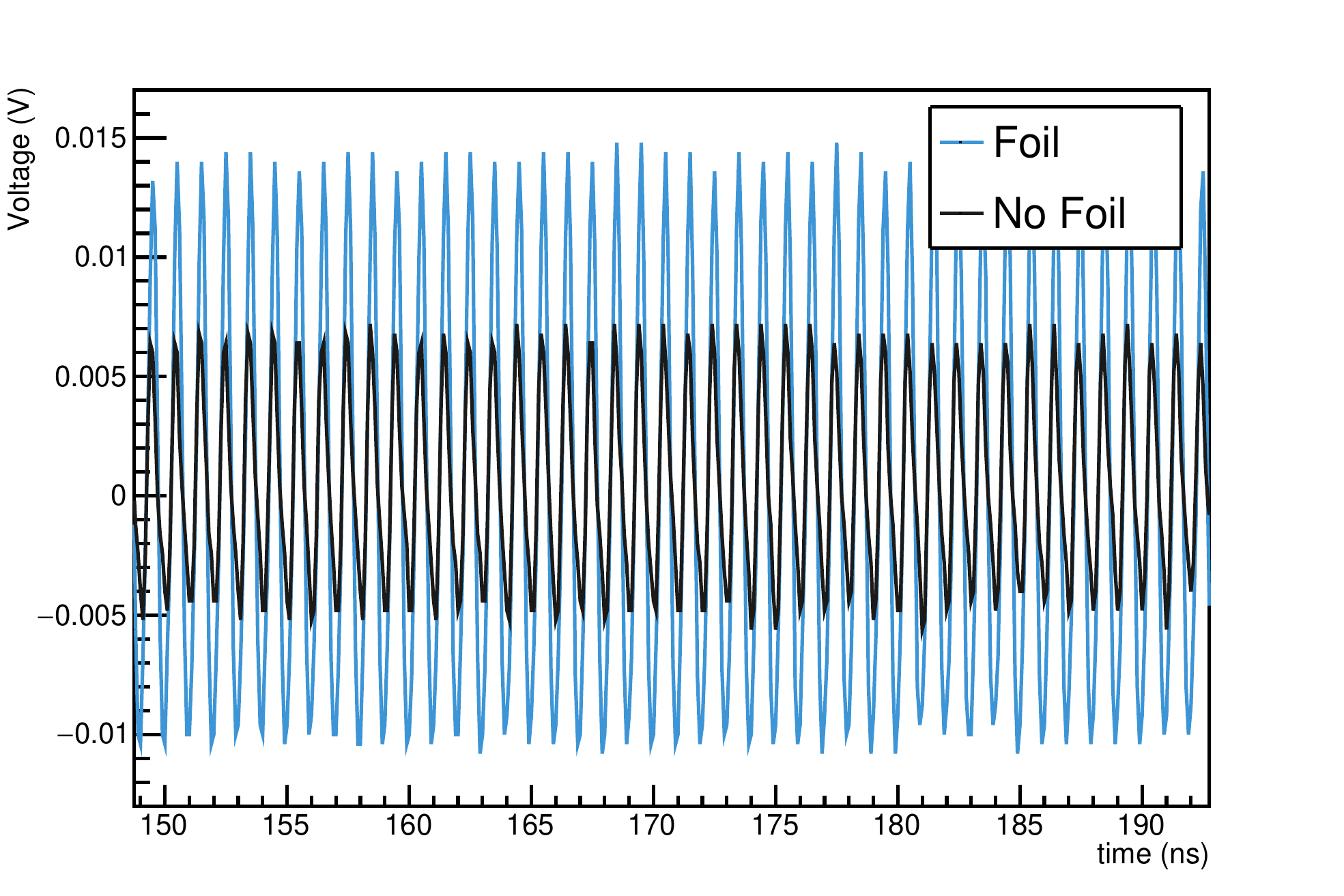}
\par\end{centering}
\caption{A `foil test', for which a conductor (here 0.3~m $\times$ 1~m) is placed at the expected ionization maximum point on the target, to estimate how a reflection should be observed in the receiver. In this test, the foil reflection is in phase with the ambient background, indicating a good receiver position.}
\label{foil_no_foil}
\end{figure}

The ESA is characterized by many sharp angles, reinforced concrete, and randomly placed metallic equipment accumulated over decades of previous experimentation. For most particle physics applications, this is irrelevant, but for radio, each conducting surface is a reflector that can affect the signal seen at the receiver. The reflections in the room were so pervasive that moving a receiving antenna relative to the transmitter by several centimeters could, in extreme cases, reduce the received amplitude of a CW signal by an order of magnitude. Typically such reduction is achieved through active carrier cancellation (a procedure whereby the transmitted signal is split and one half is fed directly into the line of the receiver, to be combined with the signal arriving at the receiving antenna. With proper alignment, the phase of the combined signal cancels the otherwise-large carrier in the receiver completely, and thus allow for smaller SNR signals to be seen in the receiver stream), but at ESA, reflections from myriad surfaces required scanning for receiver nulls empirically. Once a receiver was positioned at such a null, an additional `foil test' calibration was performed to verify that the addition of a reflecting surface at the expected location of the reflecting shower would result in a clear signal enhancement (compared to the no-foil configuration) at that receiver. 
For some configurations, it was observed the foil test would result in a further nulling of the signal, indicating a poor receiver location for that particular frequency. For others, such as the one shown in Figure~\ref{foil_no_foil} the amplitude of the carrier with the foil in place is approximately twice larger than without, indicating a favorable receiver position.

The foil tests were also quite useful from a simple physics standpoint -- reflections with a piece of foil on the order of the expected dimension of the ionization plasma gave a crude approximation of the signal amplitude to be expected during the run. In good agreement with pre-run simulation, amplitudes of $\cal{O}$(1~mV) were observed.

\subsection{Beam Splash}
The second, far more challenging background was the so-called $\sim$100 mV, several hundred nanosecond duration `beam splash', which likely is the result of somewhat complicated physics at the point at which the beam strikes the target. This background likely combines sudden appearance~\cite{sudden_appearance}, transition radiation~\cite{gorham_tr,krijn_eas,tr_motloch}, and Askaryan radiation\cite{askaryan_orig, askaryan}, plus myriad reflections from the room and from within the target itself. Beam splash was observed at all values of $\theta$, as shown in Figure~\ref{ict_dist}, but was more pronounced in the forward beam direction, as expected (additional details on beam splash will be presented in the analysis section). 

It is worth noting that beam splash would not be present for an experiment seeking to use this technique to detect in-ice neutrinos. The only background to the radar signal, from the shower itself, would be the Askaryan signal, over a very restricted solid angle.


\section{Data Analysis}
This section describes data analysis for one of the cleanest runs, towards the end of the experiment, when most of the backgrounds had been at least partially-characterized. 
We follow a technique which employs several different methods of matrix decomposition to filter background and extract signal~\cite{bean_ralston}. An alternative method using the same techniques is presented in the Appendix. 
We present evidence for a possible signal, and suggest a follow-up beam test with slightly different parameters to definitively establish this signal.

\subsection{Setup}
For this run (run 11), antennas were aligned vertically (VPol), and there was no active carrier cancellation. The present analysis will focus on data taken using a transmit frequency of 1~GHz and 5--25~W output power. 
The layout of the receivers are given in Figure~\ref{setup_run_11}, and the plots to follow are based on data taken from channels 1 and 2, which were both Vivaldi receiver antennas. There was no filtration or amplification on the input of the receivers, to avoid possible saturation effects, and to initially maximize receiver bandwidth.

\subsection{Raw Data}
\begin{figure}
\begin{centering}
\includegraphics[width=\textwidth]{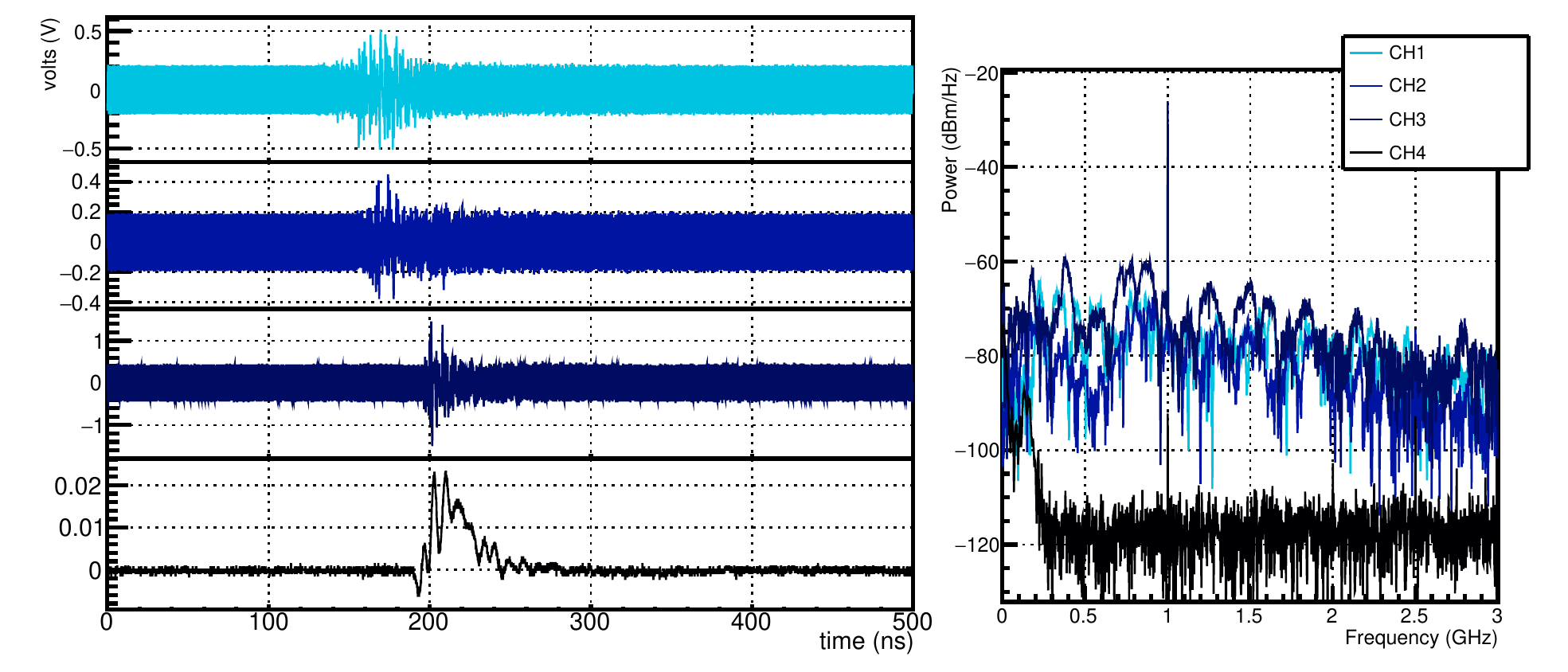}
\par\end{centering}
\caption{A typical T576 event. The 4 left panels are channels 1-4 from top to bottom, respectively, and the right panel shows their associated PSD. The offsets on the x-axis are due to air and cable propagation delays.}
\label{ex_event}
\end{figure}

Figure~\ref{ex_event} shows an event from run 11 taken in the counting house at run-time. The 4 panels on the left are the oscilloscope time traces, uncorrected for cable delays and time-of-flight. From top to bottom these are CH1, CH2, CH3, and CH4 (ICT), respectively; corresponding Power Spectral Densities (PSD) are presented in the right panel. As evident on the Figure, the amplitude of the beam splash is greater than 100~mV. In Channel~3, downstream of the beam, the amplitude exceeds 1~V. The heavy peaking in the spectrum is likely a combination of system response and the room itself, with natural nulls at certain antennas for certain frequencies, as observed during the foil tests. The carrier is evident in the PSD. 

\begin{figure}
\begin{centering}
\includegraphics[width=.9\textwidth]{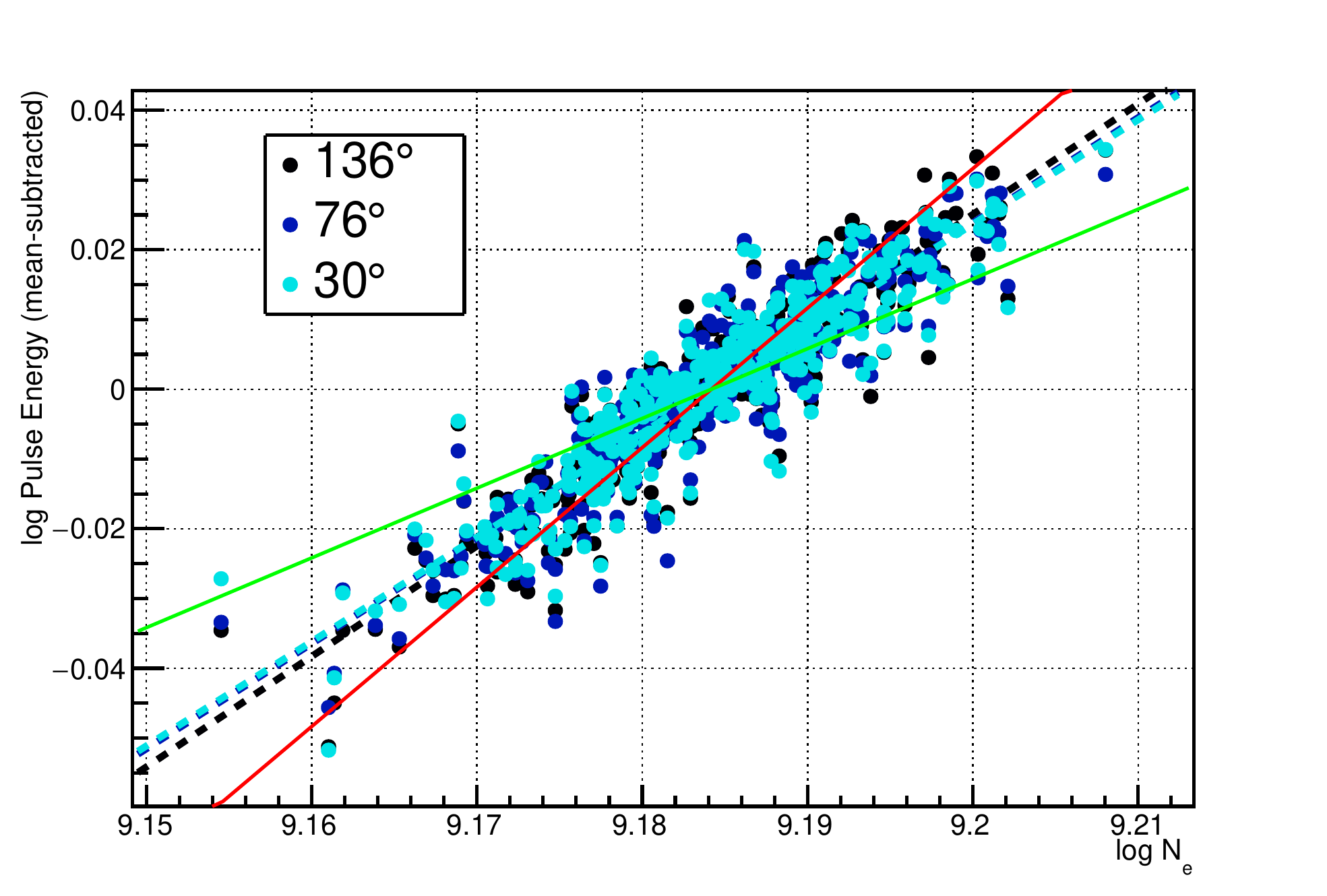}
\par\end{centering}
\caption{The ICT-measured electron number per bunch versus pulse energy, as measured by the antennas indicated by their angle from the beam momentum direction. The mean has been subtracted from all distributions in order to highlight the trend. Total coherence would correspond to a slope of 2 (red line), incoherence to a slope of 1 (green line).}
\label{ict_dist}
\end{figure}

Though the beam splash is large, it is exceedingly stable, which will later be extremely important to the background subtraction procedure. The shot-to-shot variation depends on the amount of charge in the bunch, as seen in Figure~\ref{ict_dist}, where the energy in the beam splash scales with the beam charge measured by the ICT. This is useful for building up a background `template' and for constructing `null' data, to train the analysis techniques. Previous experiments~\cite{gorham_tr} have made measurements of transition radiation which show a quadratic scaling of TR energy with beam charge, indicating coherence. The electron number $N_e$ only varied by roughly 20\% during our run, 
but our fit in log-log space has a slope roughly halfway between the expectation for complete incoherence (slope=1, corresponding to the green line in Figure~\ref{ict_dist}) and complete coherence (slope=2, corresponding to the red line). Interestingly, as the receiver is moved relative to the shower, the coherent contribution of the beam splash increases, albeit only slightly. 
To improve our signal sensitivity, the data at this point are up-sampled 
by a factor of 5 and then filtered at $\pm$300~MHz from the carrier using a time-domain software bandpass filter.

\subsection{`Null' Data}

To train the background-subtraction procedure, we developed a routine for building up what will be called `null' data, which are devoid of signal. These consist of carrier-only (beam OFF) data events added to beam splash-only (carrier OFF) data events, and constructed as follows:

\begin{enumerate}
\item A real event is selected from the data file.
\item A carrier-only event (beam OFF, carrier ON) is selected from a carrier-only file with the same frequency and output power settings. It is matched to the real event in both amplitude (via scaling) and phase (via cross correlation with the first 100~ns of the real event).
\item 
A template of beam-only events is produced by averaging over a beam-only run of 90 events. This template is then scaled using the measured value from the ICT and aligned, in time, with the real event via cross correlation, windowed around the beam onset.
\item Now that the carrier-only event is aligned with the real carrier in the pre-signal region, and the beam-only event is aligned with the real beam splash in the signal region, the carrier-only and beam-only events are summed together to produce a `null' event, which contains no signal.
\end{enumerate}

\begin{figure}
\begin{centering}
\includegraphics[width=.7\textwidth]{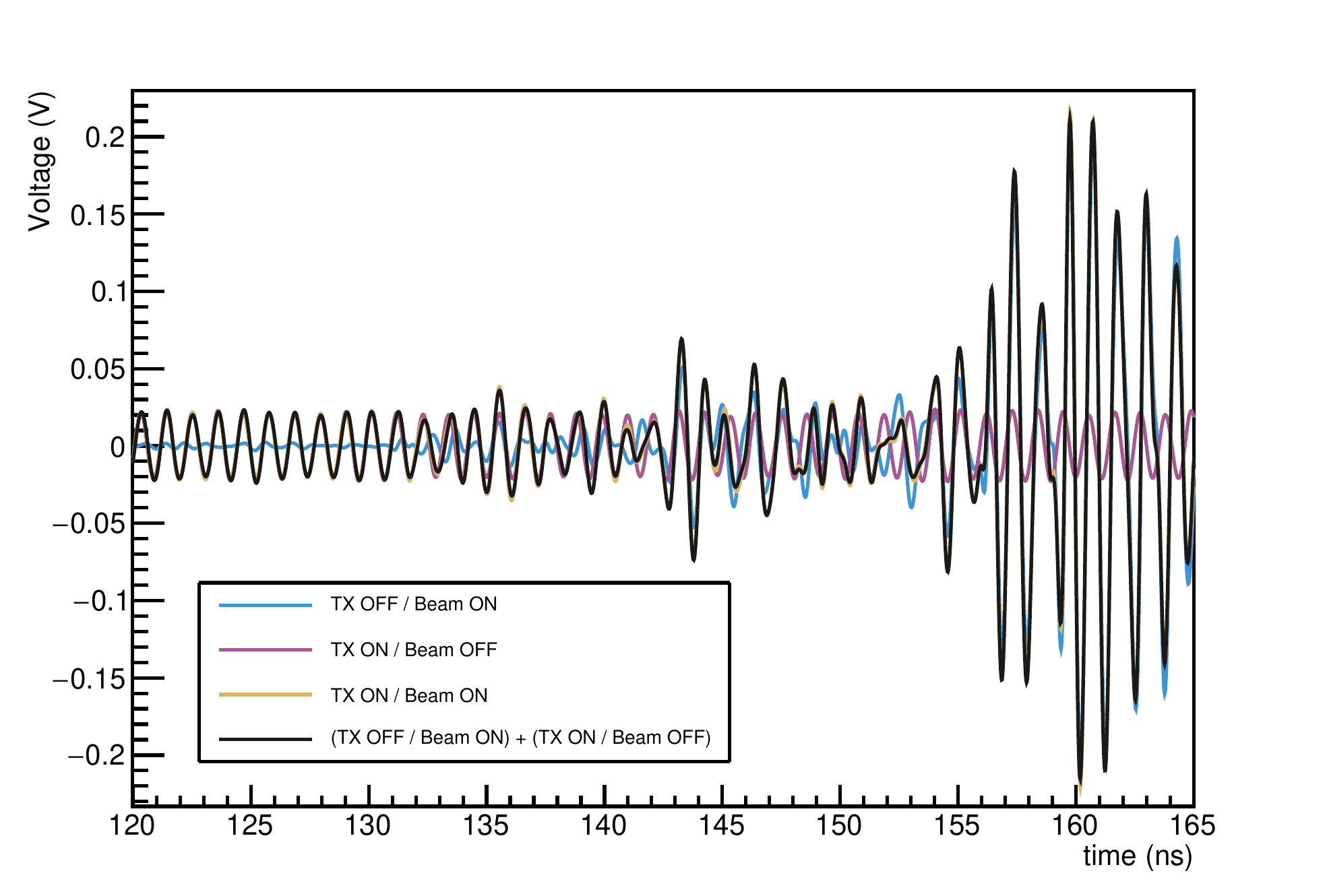}
\par\end{centering}
\caption{An example of `null' data construction. A carrier-only (TX ON / Beam OFF) event is phase aligned with the carrier of a real event, and a beam-only (TX OFF/ Beam ON) event is phase aligned with the beam splash of a real event. The two are summed to produce an event which mimics real events in phase and amplitude.}
\label{real_v_null}
\end{figure}

An example of this procedure is shown in Figure~\ref{real_v_null}. Indicated in the Figure are the carrier-only (TX ON/Beam OFF) and beam-only (TX OFF/Beam ON) events used to make the null event, along with the real event and the resultant null event. This method of construction of the null data is important because the phase relationship between the carrier and the beam splash changes from shot-to-shot, so any analysis technique needs to treat this variation carefully. It is essential that our null data set has identical carrier/beam phase relationships as the real data. Many techniques for background reduction, such as simple averaging, will fail in this case, due to the lack of fixed phase in the carrier. Similarly, monitoring for power scaling at the signal region is not possible, as sometimes the carrier and beam splash add constructively, and sometimes destructively. 

In what follows, every analysis step was carried out concurrently on two sets of data: real and null. The lack of a signal appearing in the null data gives confidence that any signal observed in the real data is not an artifact of our signal-extraction procedure.

\begin{figure}
\begin{centering}
\includegraphics[width=.7\textwidth]{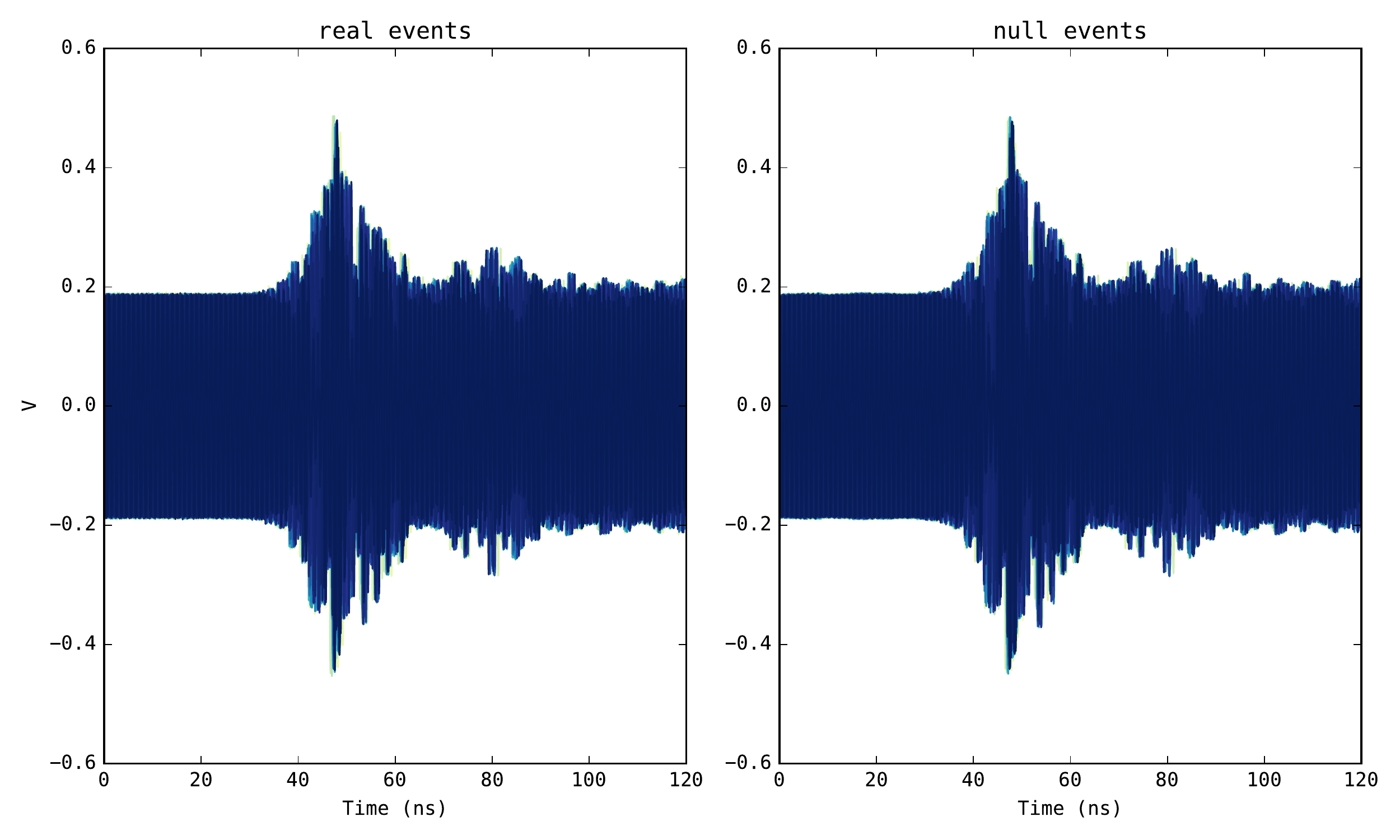}
\par\end{centering}
\caption{The data sets used in this analysis. Left: Real data from the run. Right: Null data produced via the procedure described in the text.}
\label{rf_set}
\end{figure}

\subsection{Overview of Analysis Techniques}
Figure~\ref{rf_set} shows the full data set for this analysis (e.g. all events from the cleanest run) as well as the associated null set. 
The analysis methods employed are based closely on~\cite{bean_ralston}, with necessary modifications specific to T576; we will use the vocabulary of that reference here. The procedure resembles pattern-matching routines for signal processing such as the Karhunen-Loeve technique, and is particularly suited to low-SNR data. It primarily involves decomposition of data into a basis of \textit{patterns}, which are orthogonal modes (analogous to Fourier modes) that describe the data. The power of the process, built on singular value decomposition (SVD), is that instead of pre-defined modes, as in Fourier or wavelet decomposition, the SVD method finds an orthogonal basis within the data itself to describe the data. This basis of patterns, or `eigenpatterns', or `modes' (these terms will be used interchangeably) is ordered in significance by corresponding singular values, or eigenvalues (i.e., weights). The relative scale of the weight is a measure of how well the data are described by that corresponding pattern. For T576, we expect that the beam background will occupy the most significant patterns in the decomposition, and by removing these, we can then reconstruct the reflected signal event, evident as less significant patterns in the decomposition. 

Following~\cite{bean_ralston}, we use the following terminology:

\begin{enumerate}
\item A {\it vector} $V$ is synonymous with an event captured by the DAQ.
\item A {\it pattern} is a basis mode from the decomposition, or an eigenvector, weighted by its associated eigenvalue.
\item A {\it filter} $f$ is a combination of one or more patterns which can be used to isolate, and, if desired, subtract components of the data. It can be thought of as a weighted sum over normal modes (again, the analogy to Fourier modes is useful here, in that a signal is built up of a sum of weighted normal modes). 
\end{enumerate}

The Single-Valued Decomposition is symbolically defined as:

\begin{equation}
M=u\Lambda v*,
\end{equation}
where $M$ is a matrix to be decomposed, and $u$ and $v$ are matrices containing the singular vectors of $M$. These are the patterns which describe the data in $M$, and are ordered by the matrix $\Lambda$, which has the singular values along its main diagonal. The singular values are the weights of the corresponding patterns in $u$ and $v$.

\subsection{Carrier subtraction}\label{carrier_sub}

Careful removal of the carrier from the T576 data is useful in isolating the signal. Removal of the carrier via successive sine-subtract filtration~\cite{a3_diffuse} is possible, but not problem-free. First, fitting a sine wave is susceptible to fitting errors; small errors in the subtraction can mean incomplete removal of an enormous background. Fits can be improved with pre-filtering the data, but this costs information content. Second, the amplitude of the carrier in sine-subtraction must be fixed to one value. If not, the amplitude envelope must also be extracted by a fit to data, for which the amplitude may vary. Third, the presence of harmonics requires further fitting and subtraction, each potentially removing too much or too little information, and possibly introducing artifacts. 

Using decomposition to remove the carrier solves these problems, if, for example, the modulations in amplitude are periodic or in any way repetitive, and the harmonics are stable. This is because the decomposition will yield the most significant mode of the data, which is not necessarily a Fourier mode, and may be some complicated (but correlated, e.g. from the same source) structure. It can then be removed in whole or in part by the filtration method described below.

Because a carrier naturally has some periodicity, it is useful to break the data up into bins, to see if there exists an optimal binning for background-subtraction. \message{dzb: how sensitive are you to the binning? It seems a little non-intuitive that this would be a big effect} To find such a binning, we can use a decomposition. First we construct a vector $V$ out of the \textit{pre-signal} region and partition it into bins of length $D$, then we build a matrix out of these chunks, with each chunk corresponding to a row in this matrix. 

\begin{equation}
M_{ij}=V_{(i*D+j)}
\end{equation}

\begin{figure}
\begin{centering}
\includegraphics[width=.5\textwidth]{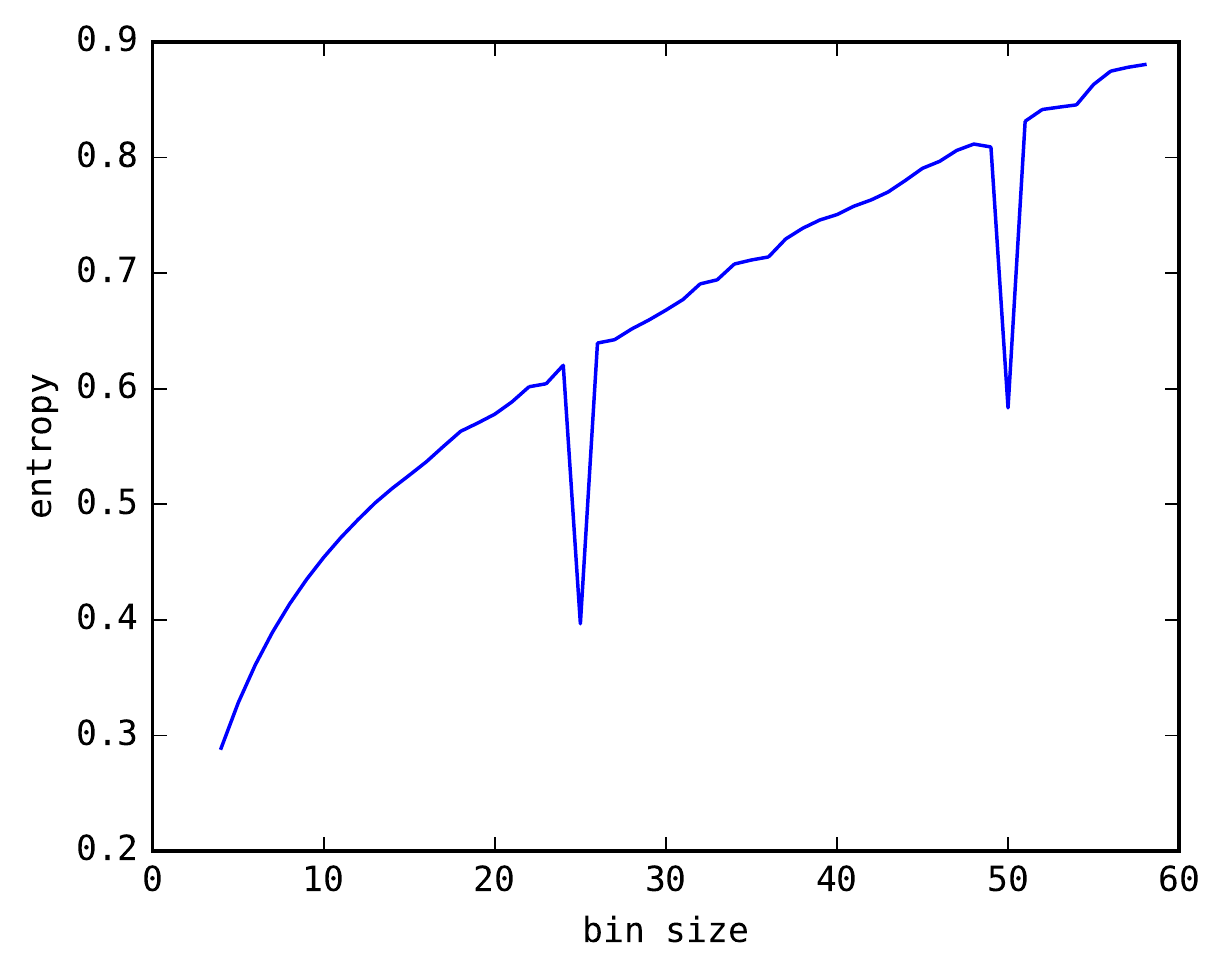}
\par\end{centering}
\caption{The Von Neumann entropy as a function of the bin size $D$, as described in the text.}
\label{entropy}
\end{figure}

If the vector has length $N$, such that there are $d=N/D$ chunks in V, then the matrix M has dimensions $d\times D$. We can then perform SVD on this matrix and examine the `orderliness' of the singular values $\Lambda$ as a function of the bin size $D$. This orderliness can be quantified by calculating the Von Neumann entropy $S=-\sum_{ij}\Lambda_{ij}\log(\Lambda_{ij})$ for the singular values in $\Lambda$ and plotting that quantity against the bin size. If there are no significant patterns (e.g. if the vector is uncorrelated noise) the entropy will exhibit the standard logarithmic dependence, and vary as $\log(D)$. A downward excursion from the $\log(D)$ curve indicates that the data is more orderly with that binning. Figure~\ref{entropy} shows such a plot for this analysis. As evident in the Figure, the values are all well below $\log(D)$, with a strong downgoing excursion at $D=25$, indicating an optimal binning for this run. 

\begin{figure}
\begin{centering}
\includegraphics[width=.7\textwidth]{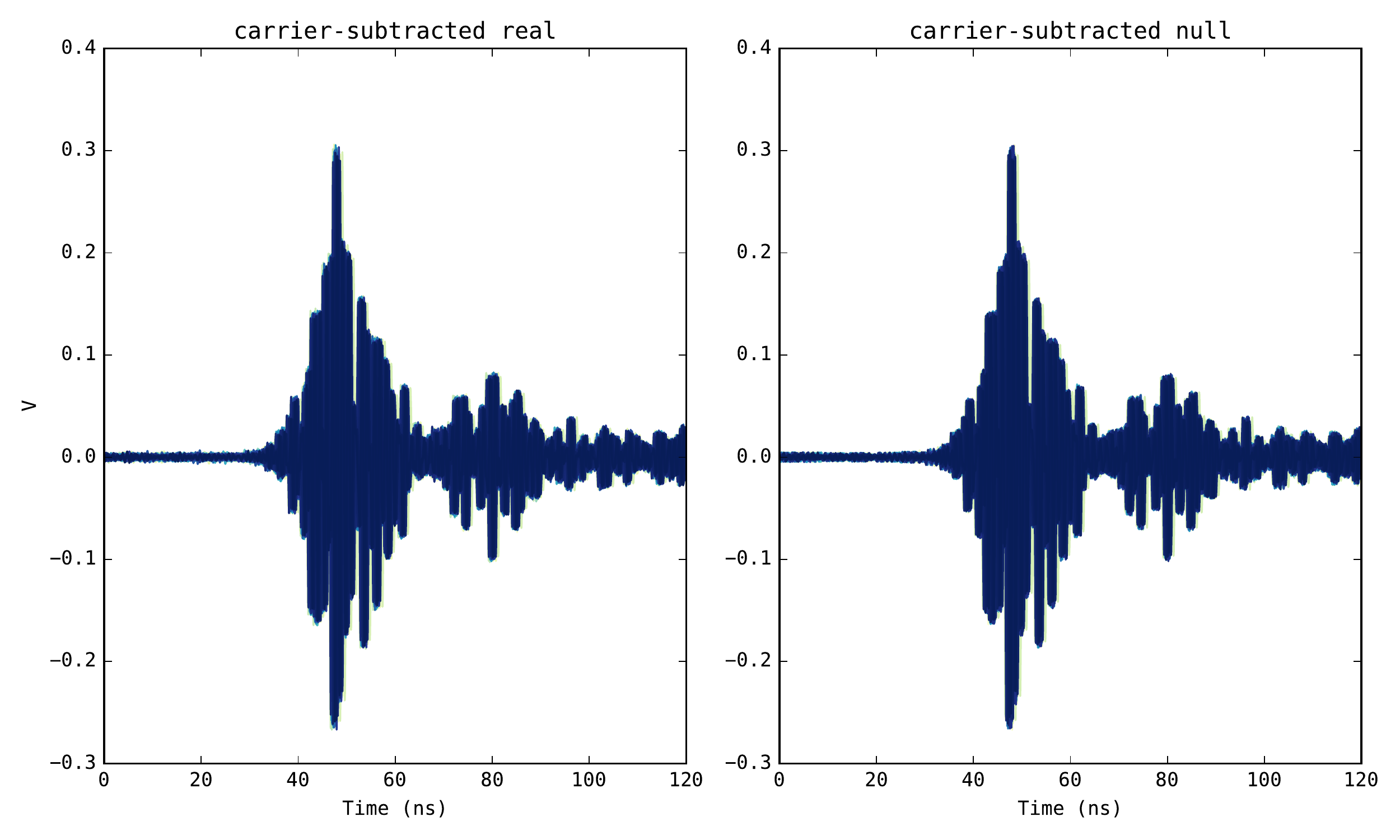}
\par\end{centering}
\caption{The data sets with the carrier removed through the process described in the text. Left: Real data. Right: Null data.} 
\label{carrier_subtract}
\end{figure}

When we subsequently bin with $D=25$, the resultant distribution of singular values indicates that the carrier can be described fully by a small number of modes $n$. We can then zero out the remainder of the singular values, 

\begin{equation}
\Lambda^\prime_{ij}=\Lambda_{ij}\Theta(n-i),
\end{equation} 
and reconstruct the matrix $M^\prime$ by reversing the SVD, using the truncated matrix of singular values $\Lambda^\prime$. The indices of the new matrix can then be flattened to recover a filter $f_{carrier}$ which can then be subtracted from the signal region of the same event. The results of this filtration are shown in Figure~\ref{carrier_subtract} for both real and null data, in which the $\sim$200~mV carrier has been reduced to the level of noise by this procedure. The carrier removal was performed on each event individually using the above procedure.

\subsection{Alignment of the Sets}

\begin{figure}
\begin{centering}
\includegraphics[width=.7\textwidth]{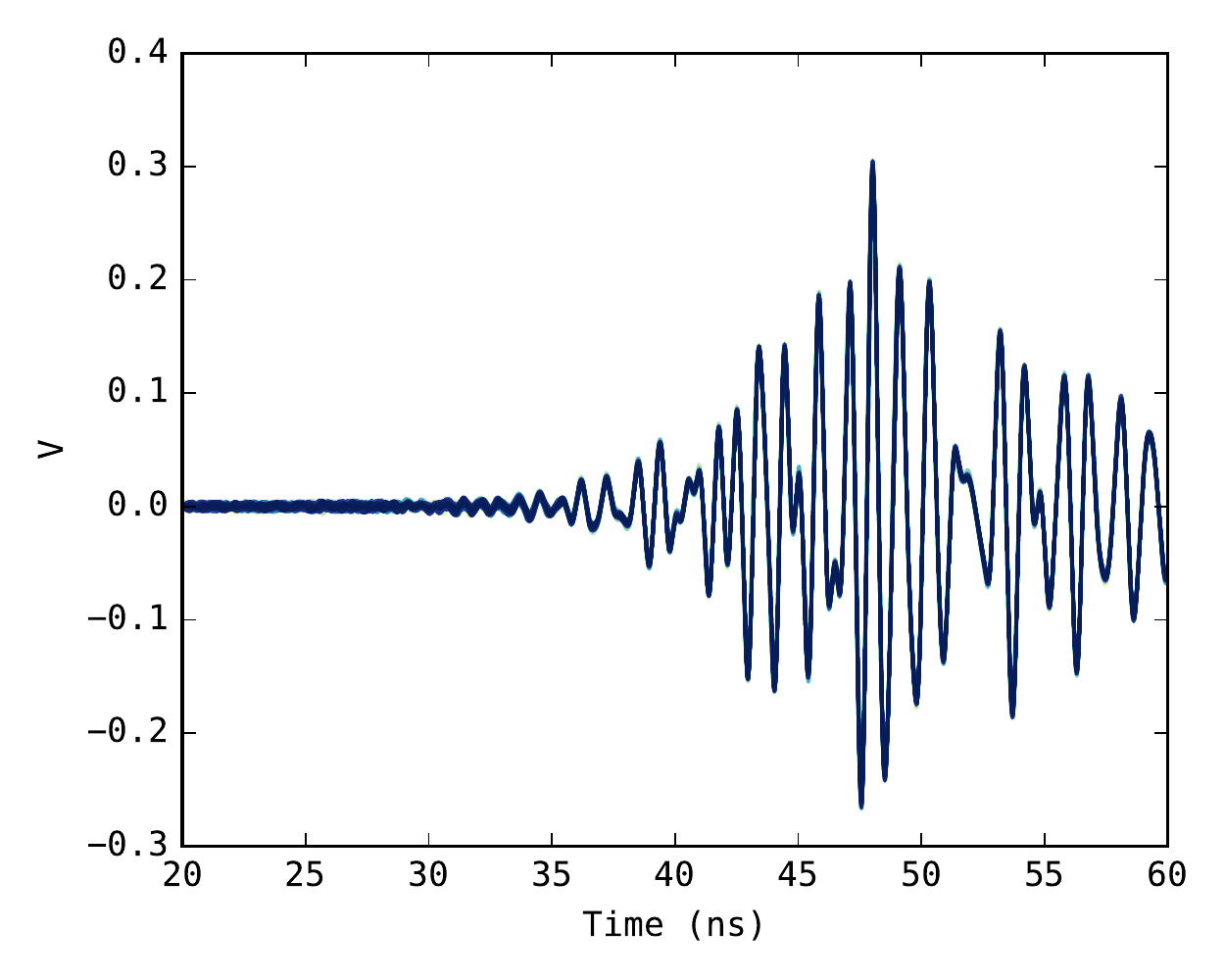}
\par\end{centering}
\caption{Carrier-removed data aligned via cross-correlation. All events from both real and null sets are overlaid ($\sim$180 events).}
\label{aligned_data}
\end{figure}

At this point we have two carrier subtracted sets, real and null, although some trigger-point jitter from the experiment still remains. We therefore subsequently align all of the events in both the real and null data sets, to the same, single reference event (selected arbitrarily, and subsequently discarded from the analysis) via a cross-correlation routine. 
The aligned, carrier-subtracted waveforms (all overlaid) are shown in Figure~\ref{aligned_data}, zoomed slightly to better illustrate the quality of the alignment.

\subsection{Extraction of the Signal}\label{extraction}

Now that we have carrier-removed and aligned waveforms, we perform decomposition of the remaining waveforms, removing the most prominent modes, corresponding to the beam splash, and the least prominent modes, corresponding to uncorrelated noise, and then, finally, performing a careful average on what remains to accentuate any possible signal.

We first build up two matrices, $M^{R}$ and $M^{F}$ for real and null data respectively, which have the following structure:

\begin{equation}
  M_{ki}=V_i^k.
\end{equation}

\begin{figure}
\begin{centering}
\includegraphics[width=.7\textwidth]{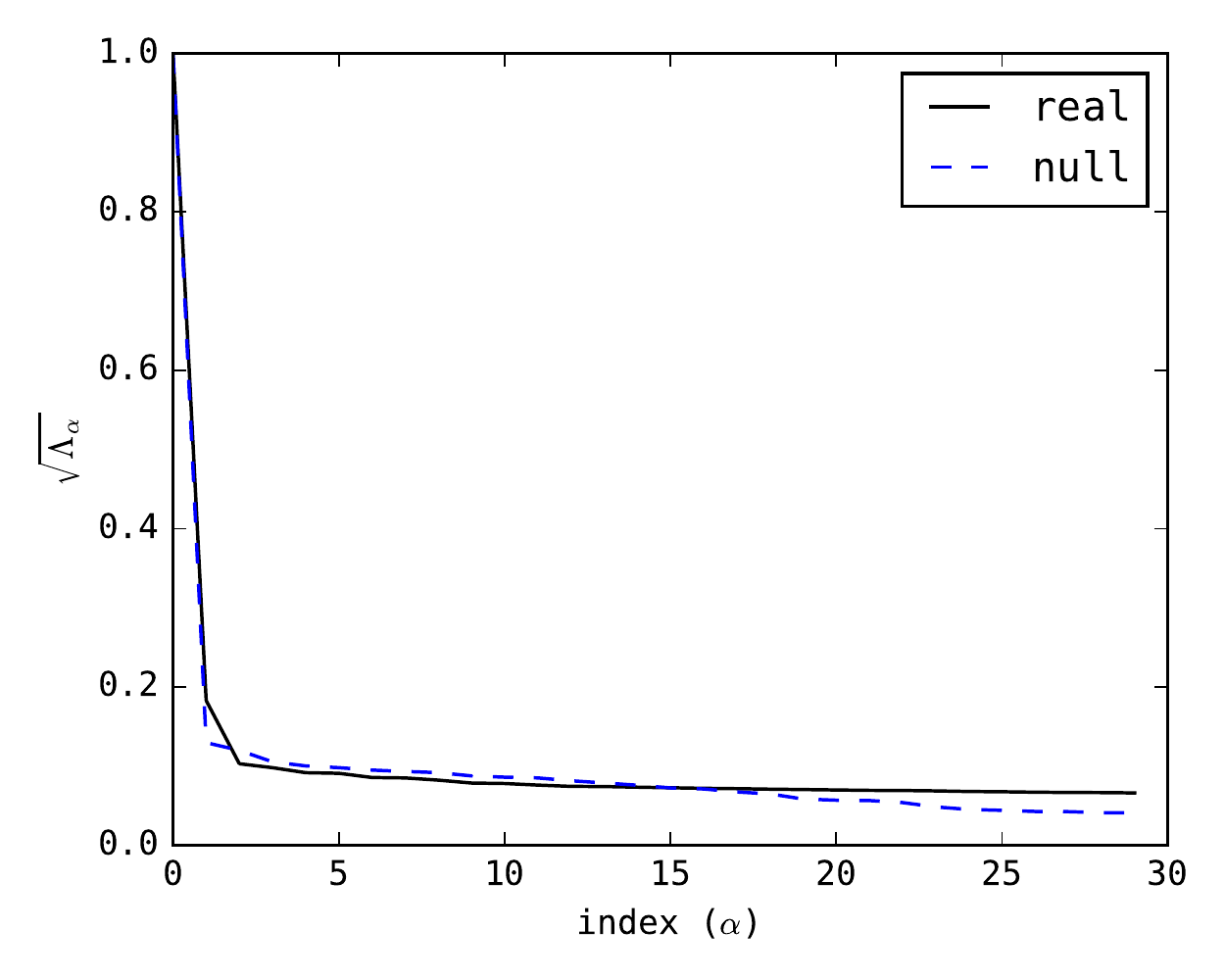}
\par\end{centering}
\caption{The normalized distributions of singular values for real and null sets after decomposition.}
\label{singular_vals}
\end{figure}

Here $k$ is a label for identifying the event (e.g. $k=1, 2...N$ for N events) and $i$ is the index of the data within $V^k$. Therefore, $M$ is a matrix in which each row is an event. We next make a decomposition of each matrix, which will simultaneously decompose all events into a basis for each full set, real and null. We then examine the normalized distribution of singular values $\Lambda_\alpha$, equivalent to the diagonal vector of the singular values within the matrix $\Lambda$. As shown in Figure~\ref{singular_vals}, where we have plotted $\sqrt{\Lambda_\alpha}$ to emphasize the shape of the curve, the two sets follow the same trend (the distribution is truncated above n=30 for readability). 

We now present the results of this analysis. The time-versus-frequency plots (spectrograms) that follow have units of power in $V^2Hz^{-1}$. The colorbar scale is the same for real and null, and varies from plot-pair to plot-pair. 

\begin{figure}
\begin{centering}
\includegraphics[width=.9\textwidth]{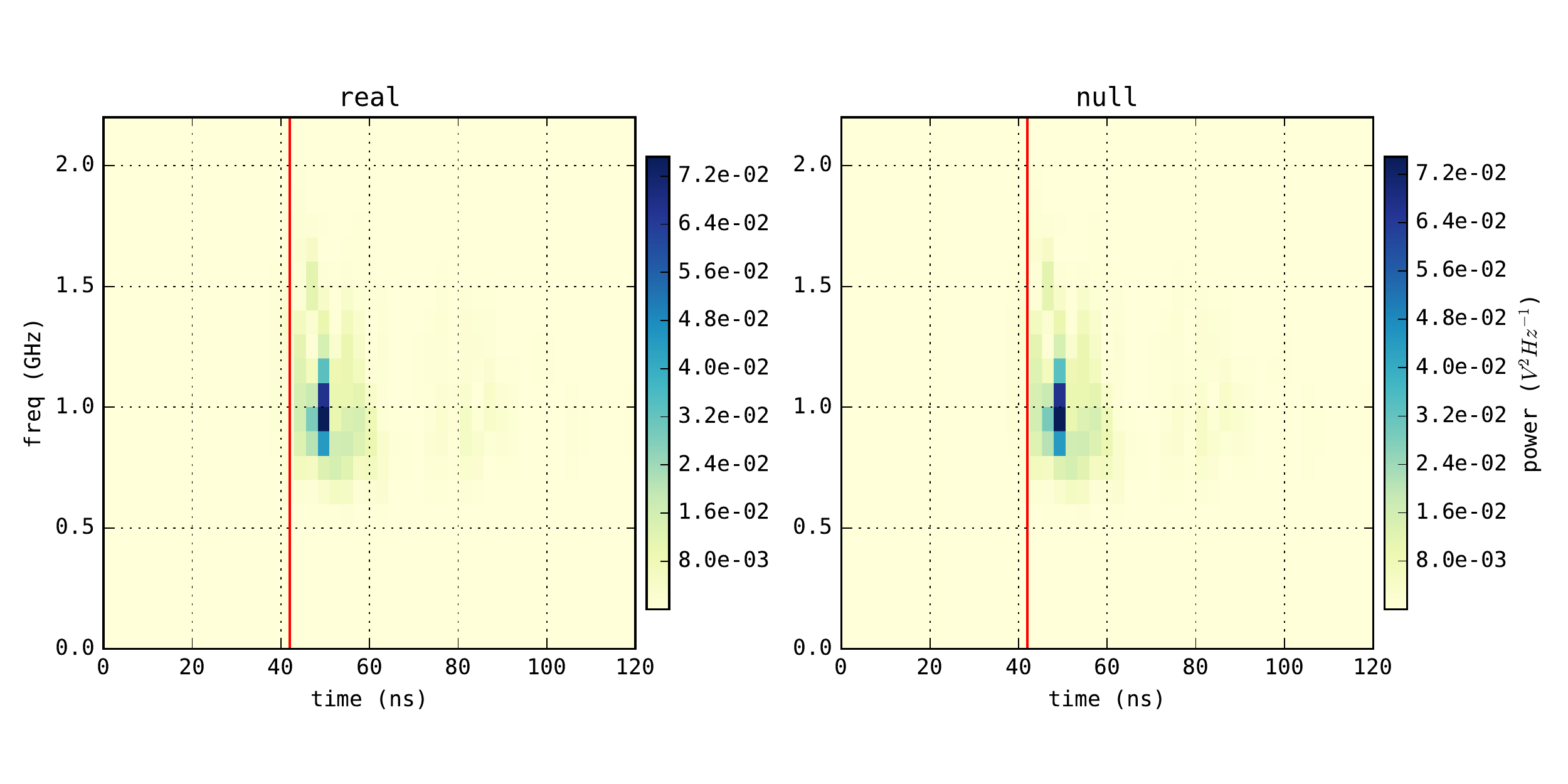}
\par\end{centering}
\caption{The average of all events, real and null, after reversing the decomposition, without removing any patterns. The solid vertical line indicates the approximate expected signal onset point, in time.}
\label{full}
\end{figure}

\begin{figure}
\begin{centering}
\includegraphics[width=.9\textwidth]{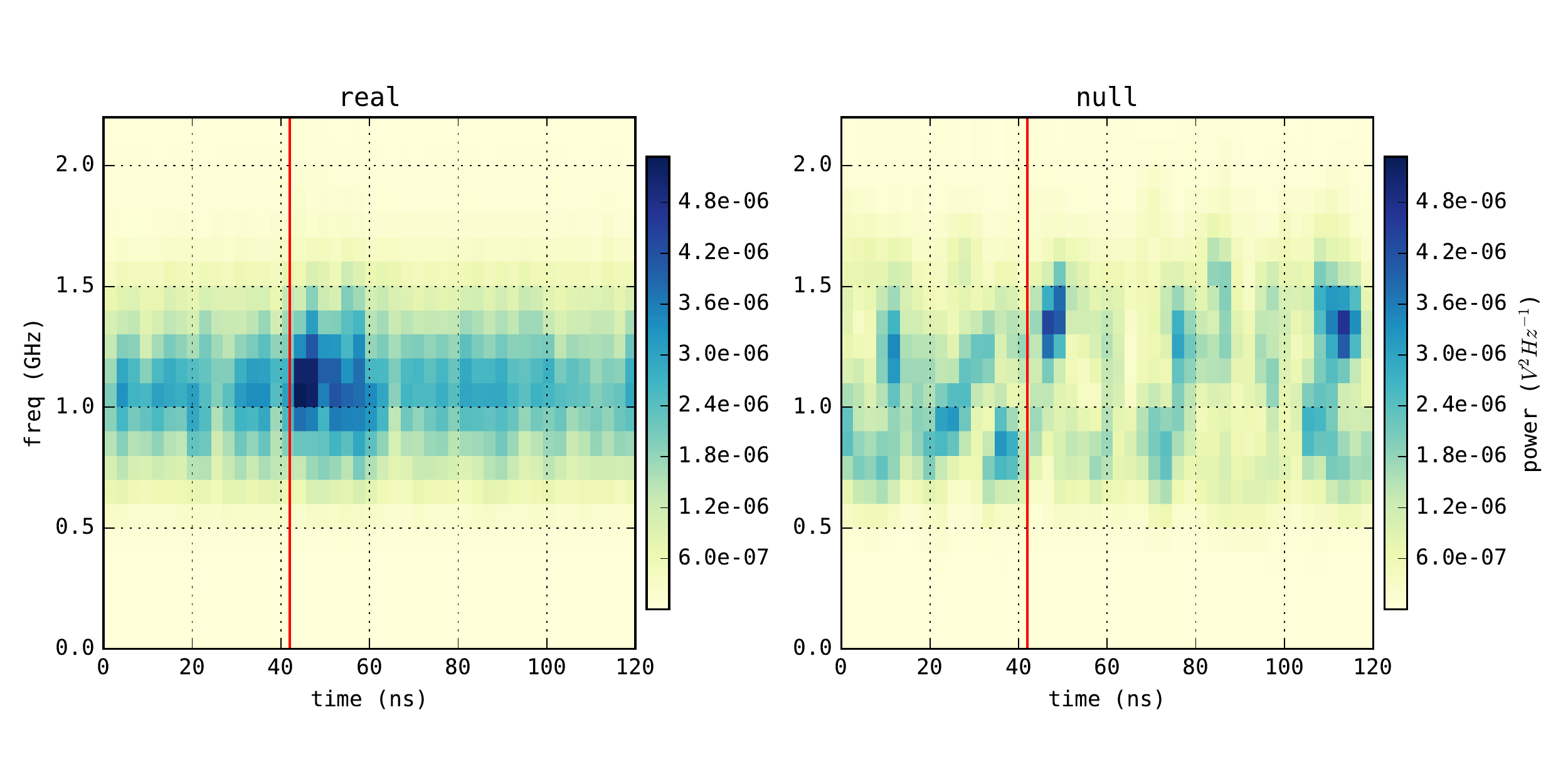}
\par\end{centering}
\caption{The average of all events, real and null, after reversing the decomposition and removing the most significant patterns, which correspond to the beam splash. The solid vertical line indicates the approximate signal onset point. }
\label{11_5_ch1_null}
\end{figure}

By inspection (via comparison to the original waveforms), it is evident that the beam splash, as expected, corresponds to the first singular values in both real and null sets. This is shown in Figure~\ref{full}, for which we plot the average spectrogram of all events in the set, real and null, after reversing the decomposition, but prior to removal of any patterns (hence $M^{R\prime}=M^R$, $M^{F\prime}=M^F$). This is useful as a reference in what follows. We then truncate the singular values in the opposite way as before, that is, we zero the most significant singular values, (and also remove the n$>$40 modes to further suppress noise) and then reverse the decomposition to recover the filtered events. Finally, we construct a power spectrogram of each event, and, finally average the spectrograms. The result of this procedure is shown in Figure~\ref{11_5_ch1_null}, in which we observe a clear difference in the time-spectral content between the real and null sets. There is a clear scaling at the time when the reflected radar signal is expected (roughly 42-45~ns into the trace as pictured) in the real set, but not in the null set. Moreover, if we selectively examine each pattern and reverse the decomposition for each one singularly, in no case do we observe scaling at the signal onset point in the null data (excepting, of course, the first two beam splash patterns).

The scaling in the signal region, which is only observed in the real data, is a suggestive hint of a signal. Of particular interest is the timing of the signal onset. It is clear by comparing Figures~\ref{full} and \ref{11_5_ch1_null} that the peak strength of the scaling in the signal region of the filtered data occurs before the peak strength of the beam splash in the unfiltered data. Calculations (based on cable delays and the known time of beam-on-target) similarly predict arrival of the reflection, in the receivers, 5-10 ns before the peak of the beam splash. 

\begin{figure}
\begin{centering}
\includegraphics[width=.9\textwidth]{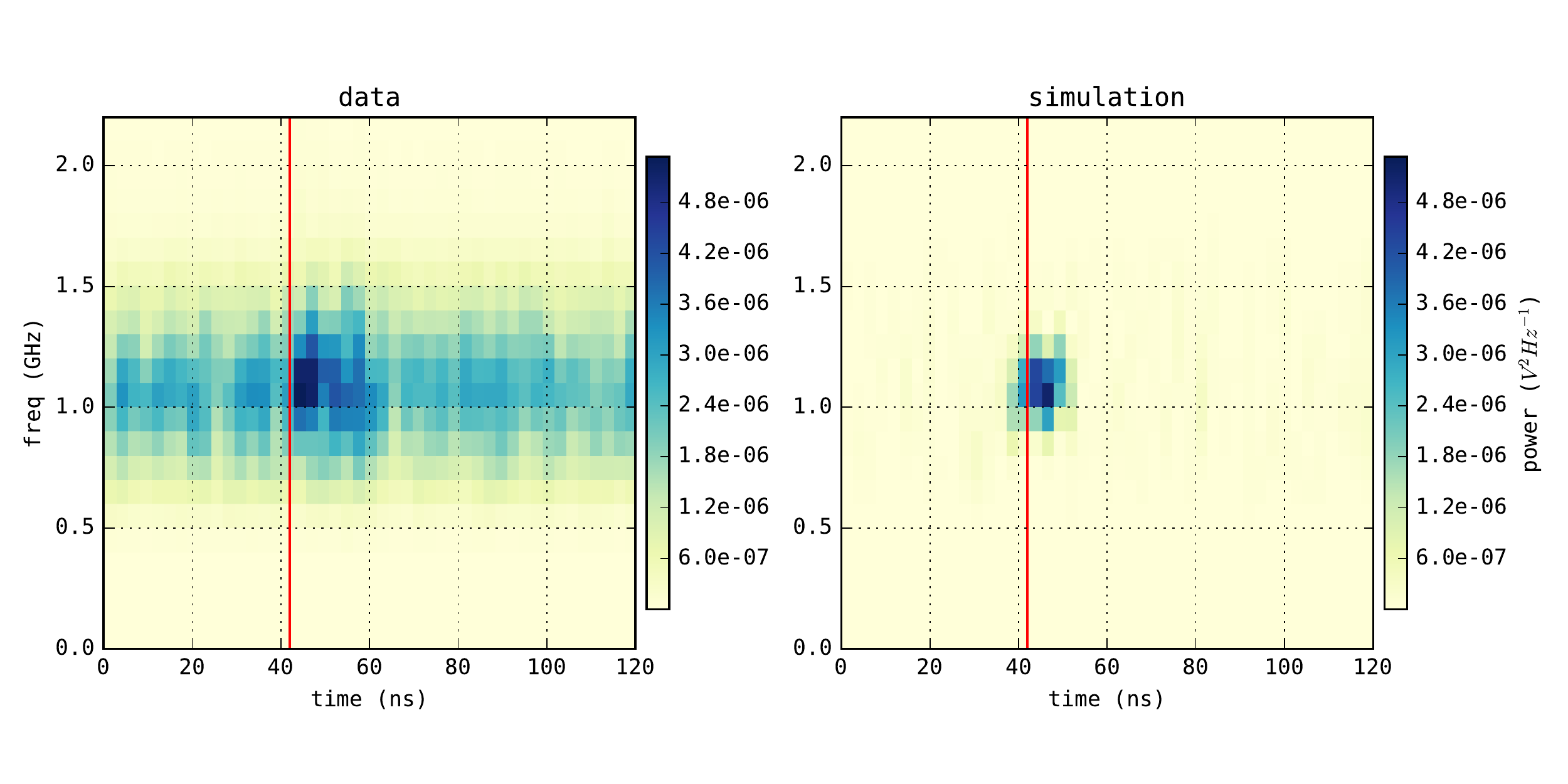}
\par\end{centering}
\caption{A comparison of the resultant filtered data to the RadioScatter simulation, for the same geometry and transmitter settings as the real run, with a plasma lifetime of 10~ns. The solid vertical line indicates the approximate signal onset point.} 
\label{sim_real}
\end{figure}

\begin{figure}
\begin{centering}
\includegraphics[width=.9\textwidth]{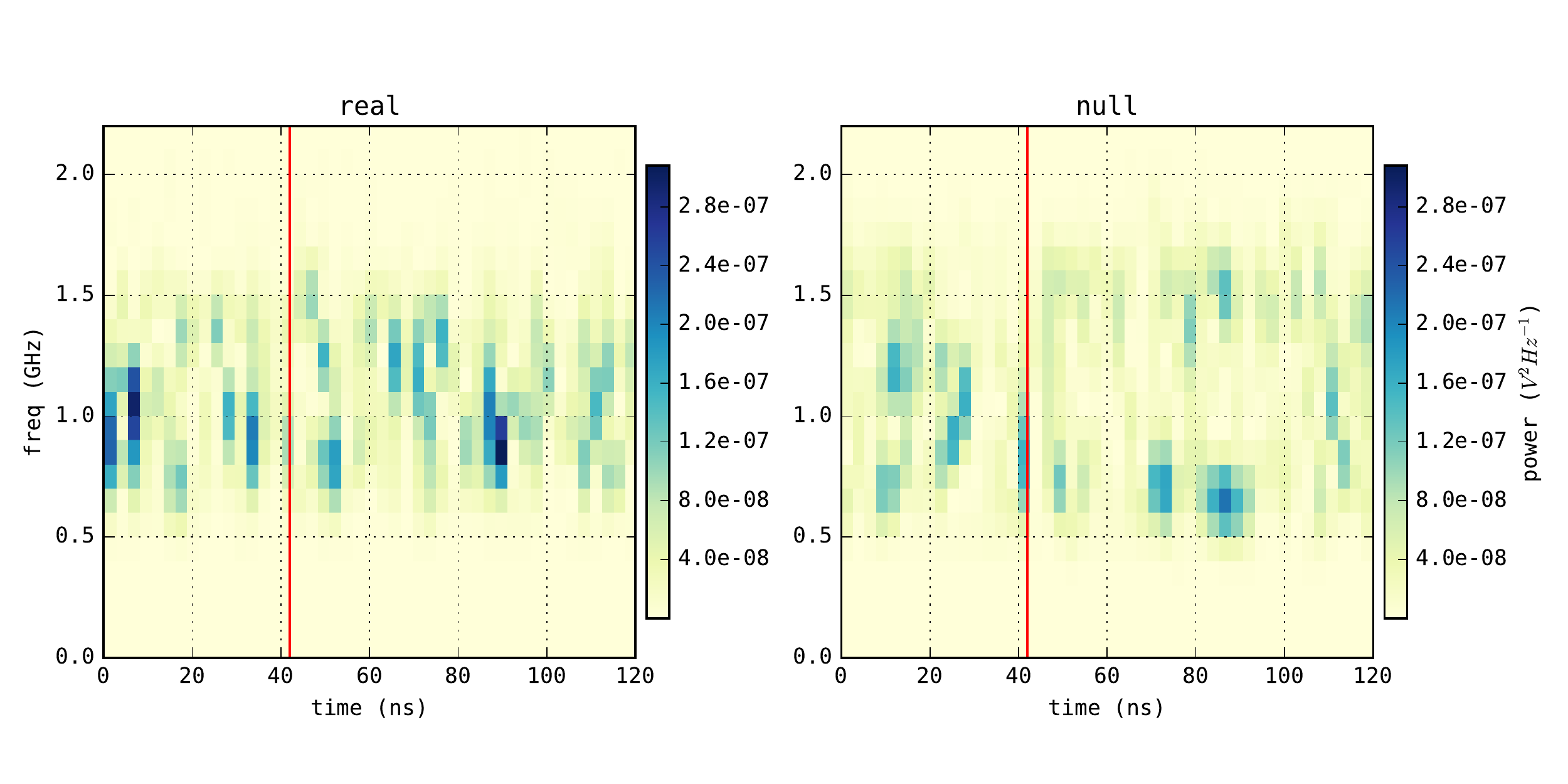}
\par\end{centering}
\caption{An example of an higher-order mode with singular value $n=15$. The solid vertical line indicates the approximate signal onset point.}
\label{noisemode}
\end{figure}

Figure~\ref{sim_real} shows a comparison of the real data to simulated signal, which has been produced using the RadioScatter code~\cite{RadioScatter} using the exact specifications of this run and a plasma lifetime of 10~ns. The agreement is quite good, with the difference in power between data and simulation less than 10\% and a very similar spectrogram shape. We note that we have not yet fully incorporated the full system response into the analysis chain; this is currently in progress. Accounting for cable losses (small, given the modest cable runs) and antenna inefficiencies are expected to reduce the signal power in the simulation by a few percent. Consistent with the noisy environment, the data trace is somewhat `messier' than the simulation. 
A more careful selection of patterns, e.g. eliminating some of the less significant noise modes, would likely clean up this background a bit. Although a full analysis of the noise modes has not yet been performed, we present an example noise mode (representative of all modes above n=10 or so) in Figure~\ref{noisemode}.

\begin{figure}
\begin{centering}
\includegraphics[width=.9\textwidth]{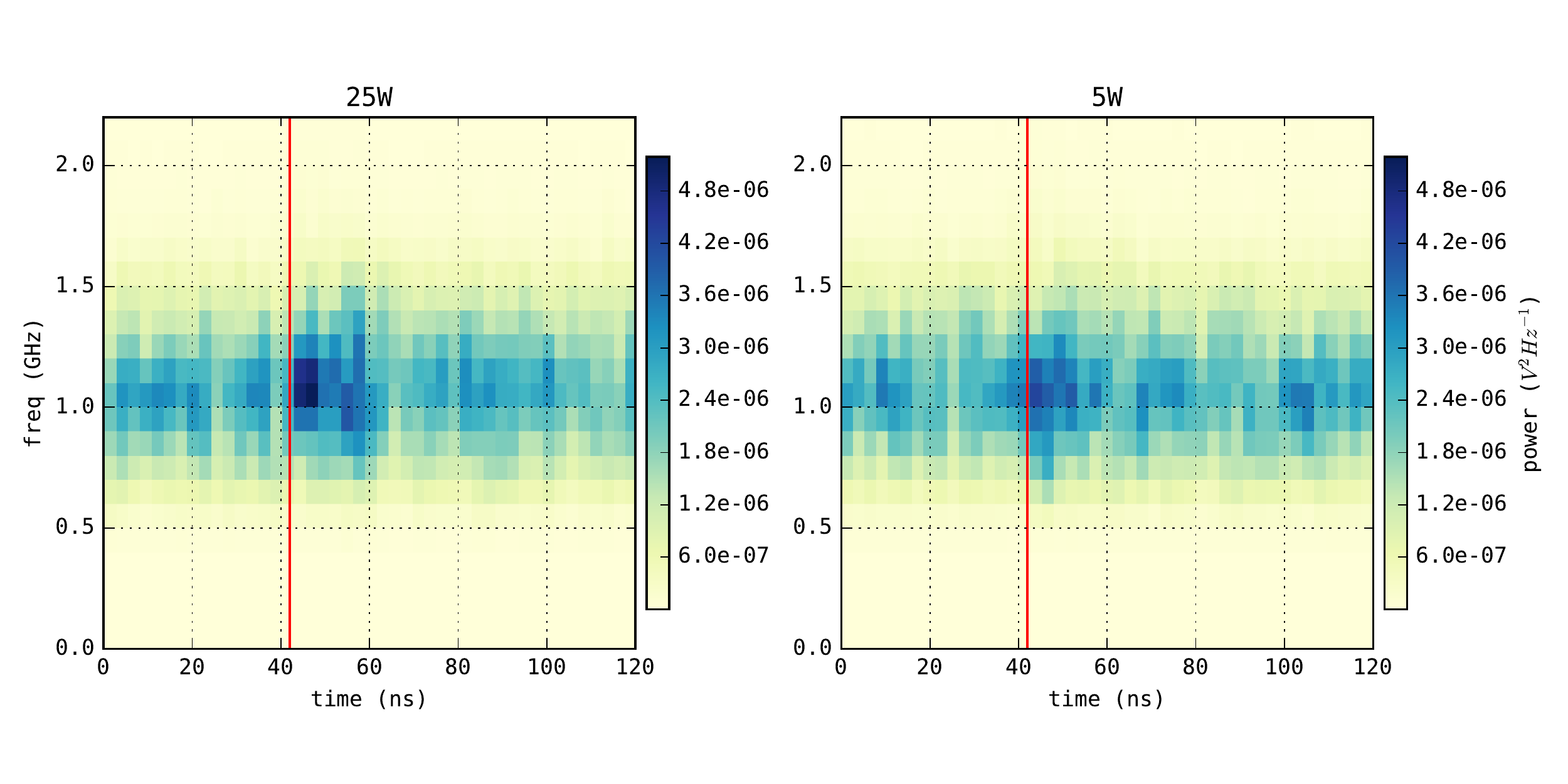}
\par\end{centering}
\caption{The average of all events, for 25~W output and 5~W output, after reversing the decomposition and removing the most significant patterns. The solid vertical line indicates the approximate signal onset point.}
\label{real_real}
\end{figure}

We follow the same procedure outlined above for data taken when the output power of the transmitter was reduced from $\sim$25~W to a nominal value of $\sim$5~W. For the latter data, the power of the carrier in the pre-signal region was actually observed to be smaller by a factor of 3.6 rather than 5 \message{I thought you said that the tx power was uncertain only to 20 percent}
The resultant summed spectrogram is shown in Figure~\ref{real_real}, where there is a suggestion of scaling in both signal regions commensurate with the different outputs, although we observe only a factor of $\sim$1.5 difference between the peak power in these two spectrograms rather than the factor of 3.6 cited above. Nevertheless, the shape of the signal is similar and the time onset identical to within one bin. 
Simulations predict that this configuration should produce radar reflections with an SNR of about two, in fair agreement with the data.
\message{dzb: is this based on the 3.6 factor or the factor of 5}

\section{Significance}

\begin{figure}
\begin{centering}
\includegraphics[width=.6\textwidth]{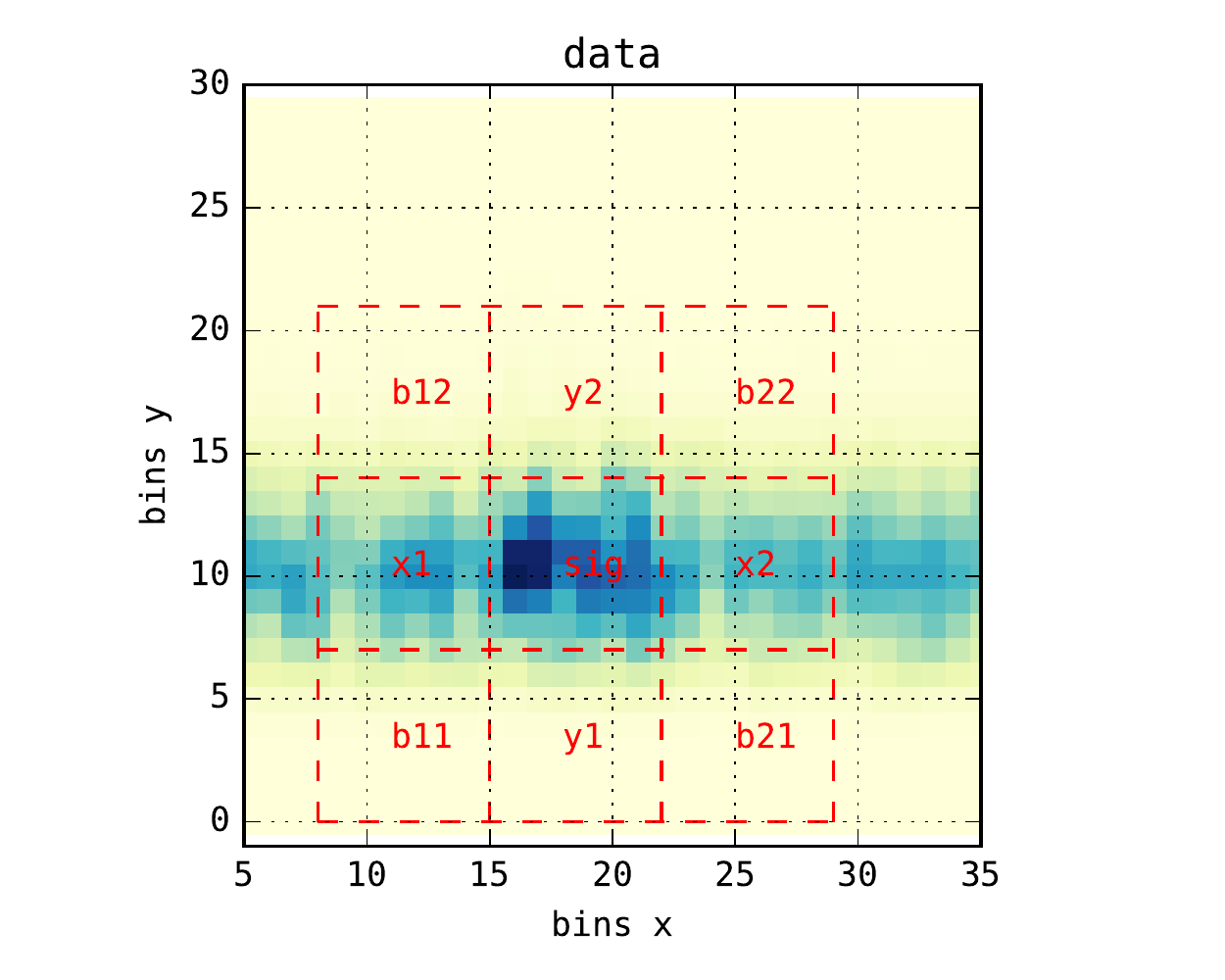}
\par\end{centering}
\caption{The sideband subtraction method. Regions with similar labels (x, y, b) are integrated, and these integrals are averaged, then subtracted from the signal region.}
\label{ss_scheme}
\end{figure}

We now present a quantitative assessment of the significance of the signal hint presented here, based on our analysis of
the real and null data sets.
To assign a significance to the observed excess, we employ a 2-d sideband subtraction technique, shown diagrammatically in Figure~\ref{ss_scheme}. The adjacent sidebands in x and y are averaged and subtracted from the signal region. This ensures that the apparent excess in the signal region is not simply a sum of the backgrounds from any residual beam splash plus carrier, at the point where they cross in the signal region. 

Mathematically, 

\begin{equation}
  <x>=\frac{\int x_1 dA + \int x_2 dA}{2},
\end{equation}
where dA indicates integration over the bounded area of the associated label, and similarly, 

\begin{equation}
  <y>=\frac{\int y_1 dA + \int y_2 dA}{2},
\end{equation}
and
\begin{equation}
  <b>=\frac{\int b_{11} dA + \int b_{12} dA + \int b_{21} dA + \int b_{22} dA}{4}.
\end{equation}
The regions `b' are selected as `ambient' background, or an overall level which sits below the beam splash remnant (`y') and the carrier-subtraction remnant (`x') such that e.g. the carrier remnant is $<x>-<b>$. The signal-region excess $\eta$ is then, 

\begin{align}\label{ssb}
\eta&=\int s dA - (<x>-<b>) - (<y>-<b>) - <b>\\
 &=\int s dA - <x> - <y> + <b>.
\end{align}

We justify this procedure as follows: if the beam splash is sufficiently broadband, as it seems to be, then the amplitude of the remainder of the beam splash in the signal region (whatever has not been fully removed by the SVD method) will be well approximated by an average of the regions co-located in time, but with frequencies above and below the signal region. Similarly, along the time axis, the remnant of the carrier (leftover from subtraction) should not prefer any particular region, therefore the average of the regions before and after the signal region should approximate the remainder of the carrier within the signal region. These backgrounds are then subtracted from the signal region. Because both of these backgrounds ($<x>$ and $<y>$) also contain an overall ambient background ($<b>$), we must add this background back in to avoid over-subtraction, as in eq.~\ref{ssb}.

\begin{figure}
\begin{centering}
\includegraphics[width=.7\textwidth]{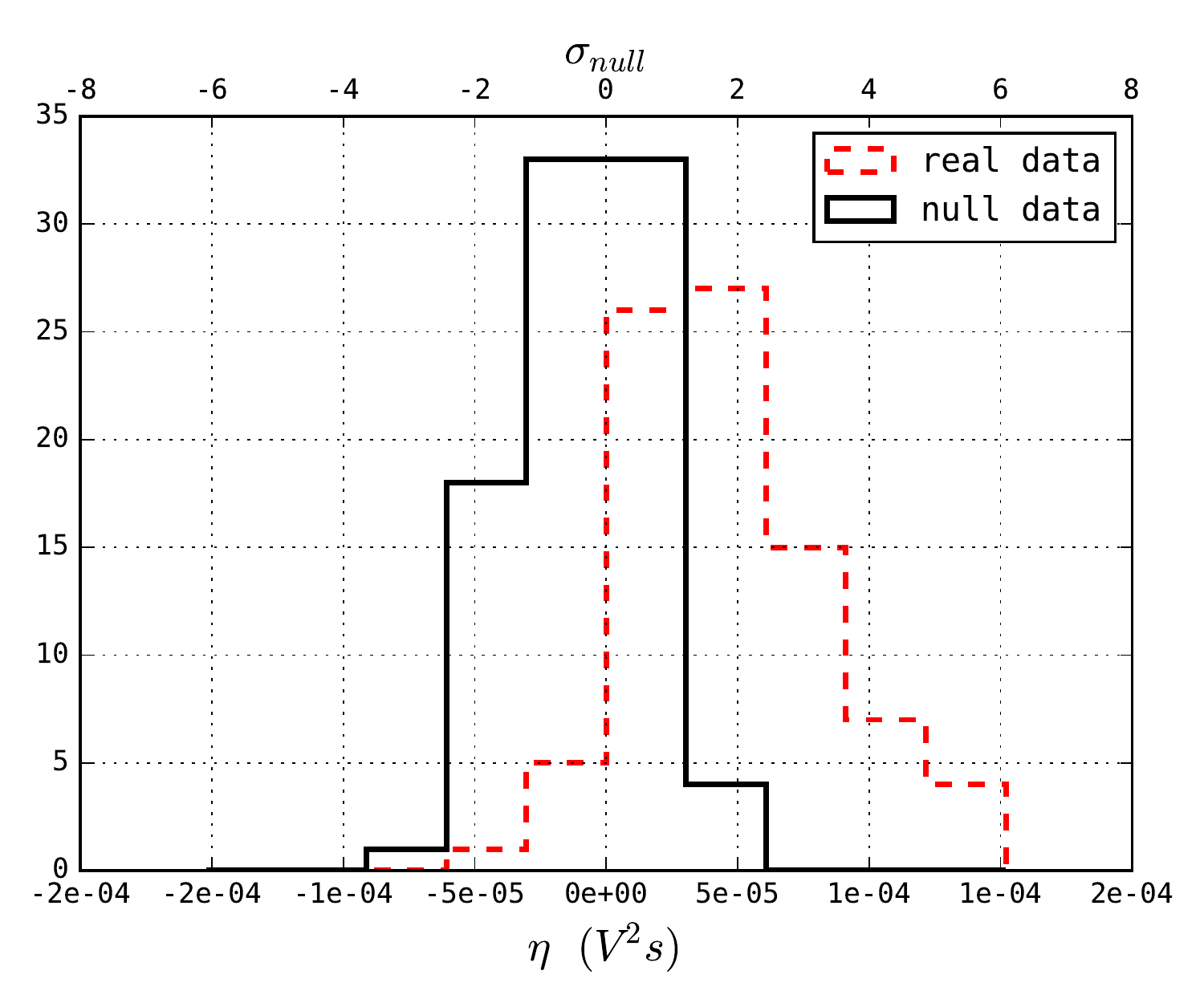}
\par\end{centering}
\caption{The significance of the signal using the 2D sideband subtraction routine. X axis units are given in $V^2 s$ and also in terms of standard deviations of the null data distribution, denoted $\sigma_{null}$.}
\label{ss_result}
\end{figure}

We perform a sideband subtraction for each event in both the data and background sets, and plot the result in Figure~\ref{ss_result}, where the x-axis is presented in units of the standard deviation of the background distribution $\sigma_{null}$. There is a clear excess in the integrated power $\eta$ for the signal events. The mean of the real data sideband-subtracted distribution is 2.36$\sigma_{null}$ from the mean of the null distribution, at -.28$\sigma_{null}$. By inspection, some of the events in the real data distribution are consistent with the null data, while some show a significance greater than 5$\sigma_{null}$. For the purposes of this analysis, we use the mean of the real data excess distribution to estimate a significance of 2.36$\sigma_{null}$.

The process of performing the same analysis on the real and null sets eliminates analysis systematics, which would be present in both. Therefore, the main systematic is in the construction of the null data.

\section{Discussion and Next Steps}

The results presented here, based on analysis of the receiver at the specular reflection point, comprise the majority of the usable data from the experiment, given the challenges outlined above. 
The signal significance of 2.36$\sigma$ is large enough to warrant further investigation. At the time of this writing, a further beam test is scheduled for fall, 2018, with several planned improvements.



In considering how this next run might improve on the results presented herein, the main challenge is clearly mitigation of the background from beam splash. 
 While some amount of RF from the shower was expected\cite{gorham_tr}, we had not anticipated such a high amplitude signal. As this is a discovery experiment, we were disinclined towards heavily filtering, so as to retain as much information as possible. However, in so doing, the large amplitude of the beam splash required a zoomed-out scope amplitude setting, resulting in only very coarse resolution given our 8-bit scope digitizer. Trigger point slewing and averaging can compensate to some degree, but, for the future, either a higher resolution DAQ or one of several known techniques to increase the digitization resolution of the data would be useful. Additionally, the most important hardware upgrade would be use of a more powerful transmitter. The simplest way to deal with the beam splash and also increase the resolution of the putative signal is simply to amplify the transmitter by another factor of 2-10, such that the reflection is well above noise after background subtraction. Noting that he beam splash amplitudes begin to fall off drastically above 1.5~GHz, we will also employ a power amplifier which has a higher frequency band of operation. 
Additionally, an option to run using a signal generator which is phase-locked to the beam arrival would be useful from an analysis standpoint. Finally, another improvement over this run will be better (or rather more well-characterized) antennas, with added directionality. This will effectively increase our transmitter power in the direction of the shower, increasing SNR. 

The main `smoking gun' signals we will look for during the next run are:

\begin{itemize}
\item a $R_1^{-2}R_2^{-2}$ dependence on the putative signal in the real data, where $R_1$ is the transmitter-shower baseline and $R_2$ is the shower-receiver baseline, 
\item a scaling of the return signal duration as a function of frequency,
\item a frequency shift of the return signal at receivers displaced from the specular point, and
\item a scaling of the return signal as a function of azimuth, with signal amplitude trending differently than the pure carrier amplitude. 
\end{itemize}

These measurements may be difficult given the restrictions of hardware and the space, but will be central to the experimental program for run 2, now that we believe our backgrounds and primary experimental challenges are well-known. 

\section{Acknowledgments}
We would like to thank the staff at the SLAC End Station Test Beam facilities and the ACR, for their invaluable hands-on support during our run, and for the consistent delivery of excellent beam and run conditions, respectively. We would like to thank the T510 collaboration for the use of their HDPE target and simulation tools for Askaryan signals. 
JT was supported by the Center for Cosmology and AstroParticle Physics at the Ohio State University. K.D. de Vries' work is supported by the Flemish foundation of scientific research (grant FWO-12L3715N). SP was funded by a US Department of Energy Office of Science Graduate Student Research (SCGSR) award. The SCGSR program is administered by the Oak Ridge Institute for Science and Education for the DOE under contract number DE-SC0014664. 


\bibliography{/home/natas/Documents/physics/tex/bib}

\appendix*
\section{Expansion of data in a basis}
We can take a data set and expand it in a basis. This basis can be a decomposition, as in the text, of any matrix. We can build up a matrix of, for example, null data, then decompose this matrix into a basis, and expand real data into this basis. This expansion can then be used as a filter for the real data. In what follows we have two data sets, a real set, and a null set. 

The general procedure is as follows: we take the null set and decompose it via singular value decomposition into a basis. We then take both sets and perform the same carrier subtraction described in Section~\ref{carrier_sub}. At this point the real set contains beam splash, whatever remains of the carrier after subtraction, and the putative signal. The null set will, by definition, contain only the beam splash and whatever remains of the carrier after subtraction.
Then, we take each real data event and expand it into the null basis. This expansion will contain only the elements in the real data which resemble the patterns in the null set, i.e., this is our filter. This expansion (filter) is then subtracted from the original event, which leaves only the components of the real event which do not resemble the null basis. 



Here we present the mathematical formulation of the basis production and data expansion. We start by building a matrix $M_{ik}=V^k_i$ where each column $k$ of the matrix $M$ is a vector from the null set $V$. We perform SVD on this matrix, 

\begin{equation}
  M=u\Lambda v*,
\end{equation}
and then systematically zero all but one of the singular values (diagonal entries of $\Lambda$), which is set to 1. 

\begin{equation}
  \Lambda^{\alpha}_{ij}=\delta_{i\alpha}
\end{equation}

For example, $\Lambda^{3}$ is a matrix with a 1 in the (3, 3) position and zeros everywhere else. Then, for each index $\alpha$, a basis vector $e^{\alpha}$ is produced by reversing the decomposition using the new matrix $\Lambda^{\alpha}$ and summing over the N columns of the reconstructed matrix $M^{\alpha}$.

\begin{equation}
  M^{\alpha}=u\Lambda^{\alpha}v*
\end{equation}

\begin{equation}\label{eq:decomp}
  e^{\alpha}_i=\sum_{k=1}^N M^{\alpha}_{ik}
\end{equation}

\begin{figure}
\begin{centering}
\includegraphics[width=.9\textwidth]{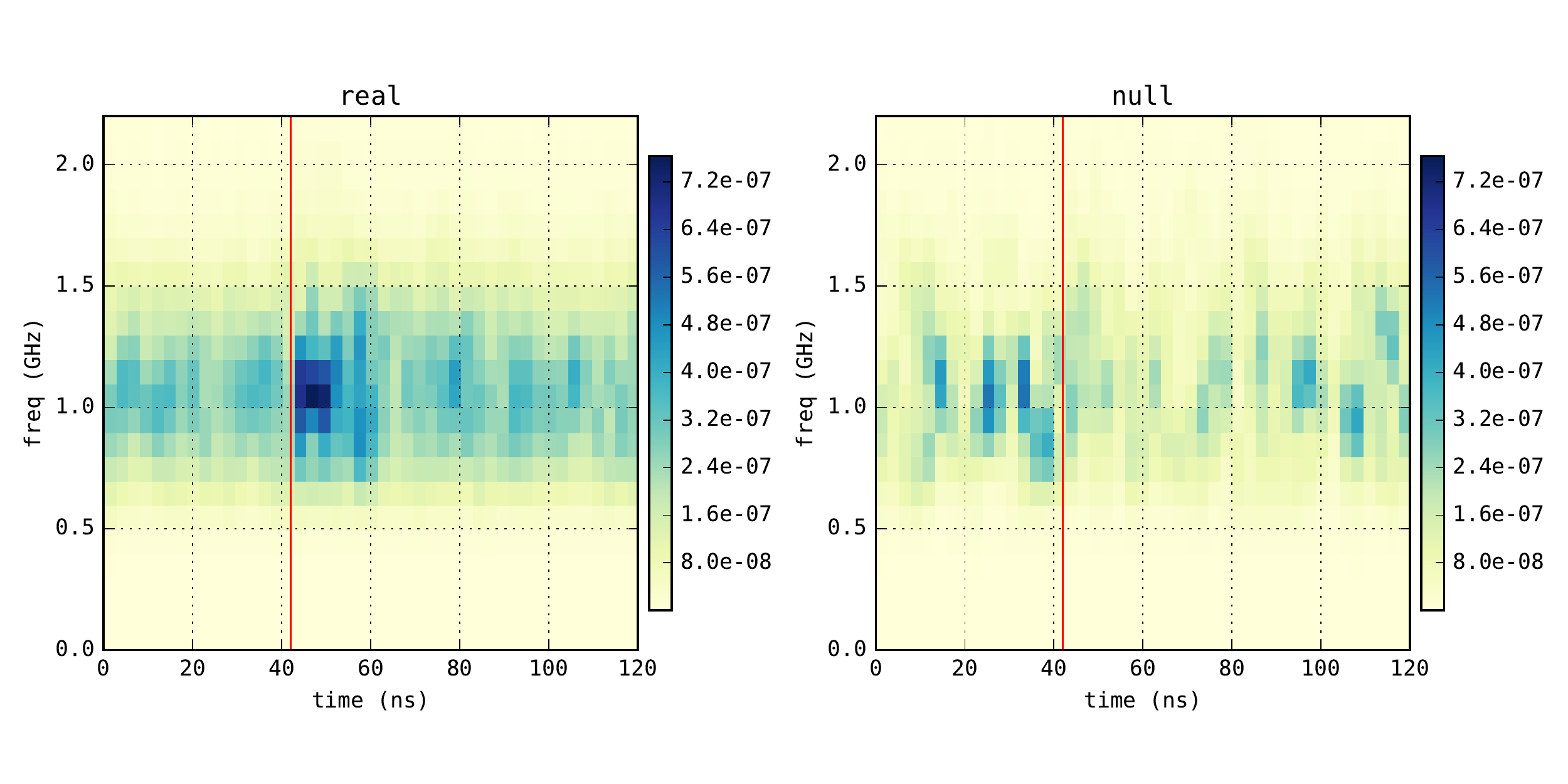}
\par\end{centering}
\caption{A comparison of the real (left) and null (right) data after the expansion in the null basis has been subtracted from each. These spectrograms are the averages of all normalized events in the sets.}
\label{expanded}
\end{figure}

We can then take a data vector $V^d$ and expand it in this basis (summation over repeated indices implied),

\begin{equation}
  c^{\alpha} = V^d_ie^\alpha_i,
\end{equation}
e.g. the expansion coefficient $c^{\alpha}$ is the inner product of the signal vector with the normalized basis vector $e^\alpha$. The expansion of the real event into the null basis is the filter $f_i=c^{\alpha} e^\alpha_i$. This filter can be subtracted from the data vector $V^d_{filtered}=V^d-f$, leaving any unfiltered excess in the data vector. For this specific case, what should remain after this filtration is the putative scattered signal, which was not present in the null set. The result of this procedure, for real and null sets, is presented in Figure~\ref{expanded}.


\begin{figure}
\begin{centering}
\includegraphics[width=.6\textwidth]{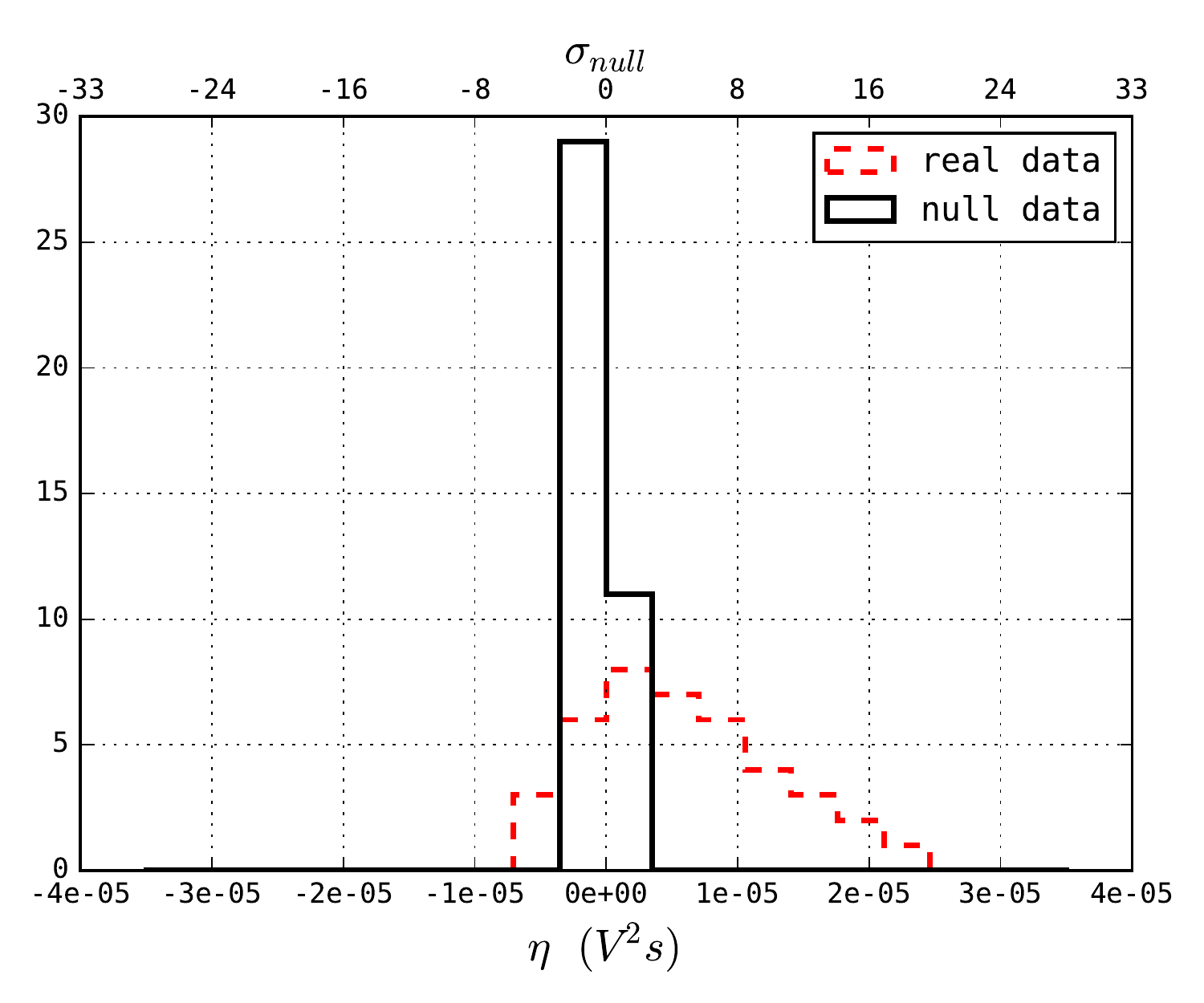}
\par\end{centering}
\caption{The significance of the excess in the signal region of Figure~\ref{expanded}, corresponding to 5.26$\sigma_{null}$.}
\label{appendix:significance}
\end{figure}

This figure is qualitatively similar to Figure~\ref{11_5_ch1_null} though the absolute amplitude is not, as here the procedure is performed on normalized vectors. The signal region excess is evident between real and null sets, which have the same color scale. Using a sideband subtraction technique (as explained in the text) we obtain the significance of Figure~\ref{appendix:significance}. 

As seen in Figure~\ref{appendix:significance}, the significance in this method is much higher than that presented in the main text. 

To investigate the effect of the basis in which the filter is constructed, we can reverse the procedure, build a basis out of the real set, and repeat the above procedure using this basis. That is, we expand real and null in the real basis, and use this expansion as a filter. This should remove everything from the null set. 
\message{dzb: I didnt understand what this means}
However, since a real scatter is phase unstable on an event-by-event case, any remainder in the signal set would possibly be indication of signal.
When building a basis, the features which are most similar event-to-event (e.g. the beam splash) will be most prominent, but phase unstable features will be diminished.\footnote{Contrast this with the method of section~\ref{extraction}, in which the overall sum was made on the spectrogram of the reconstructed events after the beam-splash patterns had been removed event-by-event. In that case, the individual phases of the signal modes did not add destructively because they were being added in power.} In this case, the production of the basis vectors averages the reconstructed vectors (see Eq~\ref{eq:decomp}), which results in destructive interference for anything which is not phase stable. 

\begin{figure}
\begin{centering}
\includegraphics[width=.9\textwidth]{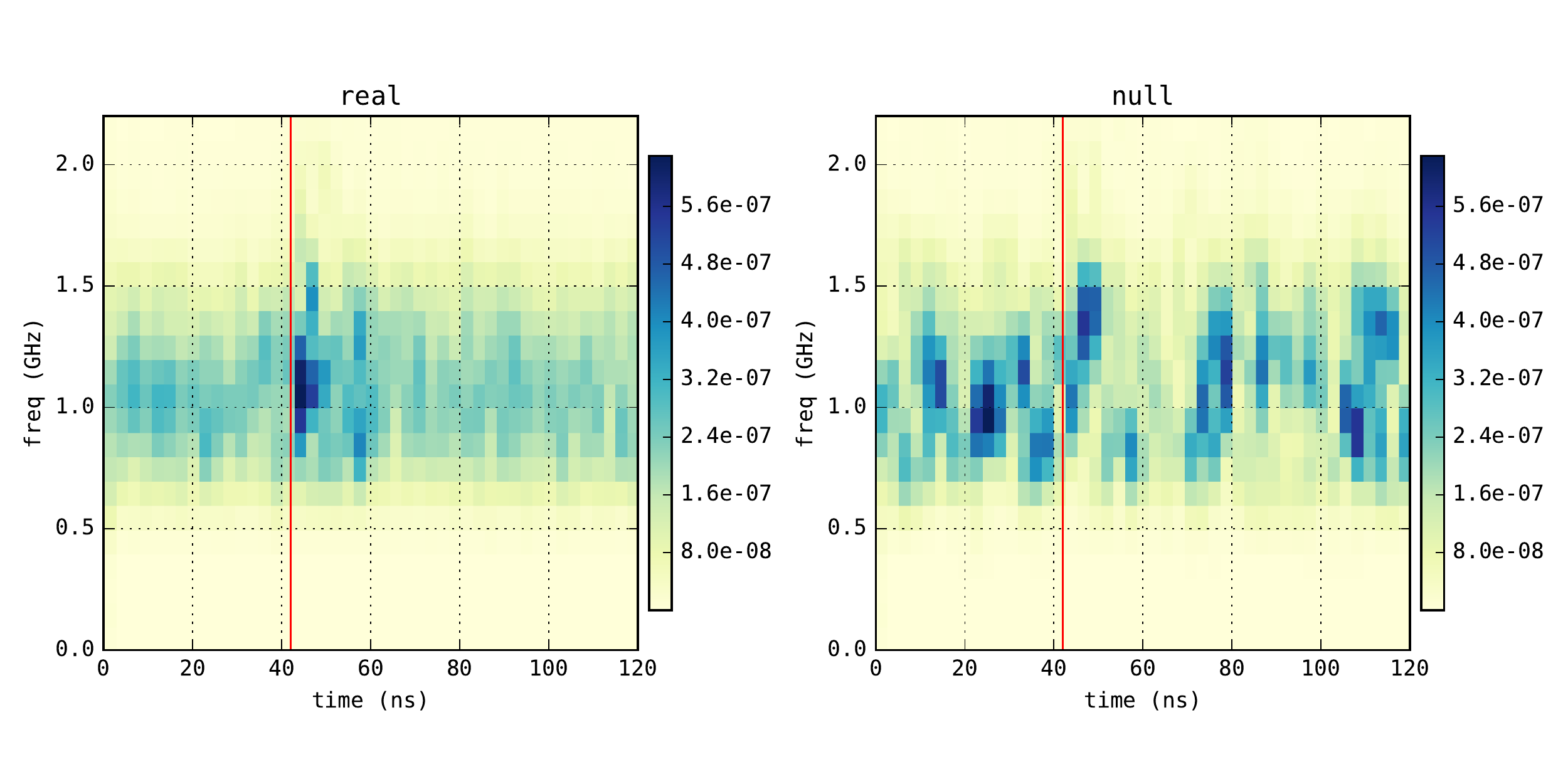}
\par\end{centering}
\caption{A comparison of the real (left) and null (right) data after the expansion in the real basis has been subtracted from each. These spectrograms are the averages of all normalized events in the sets.}
\label{expanded_2}
\end{figure}

\begin{figure}
\begin{centering}
\includegraphics[width=.6\textwidth]{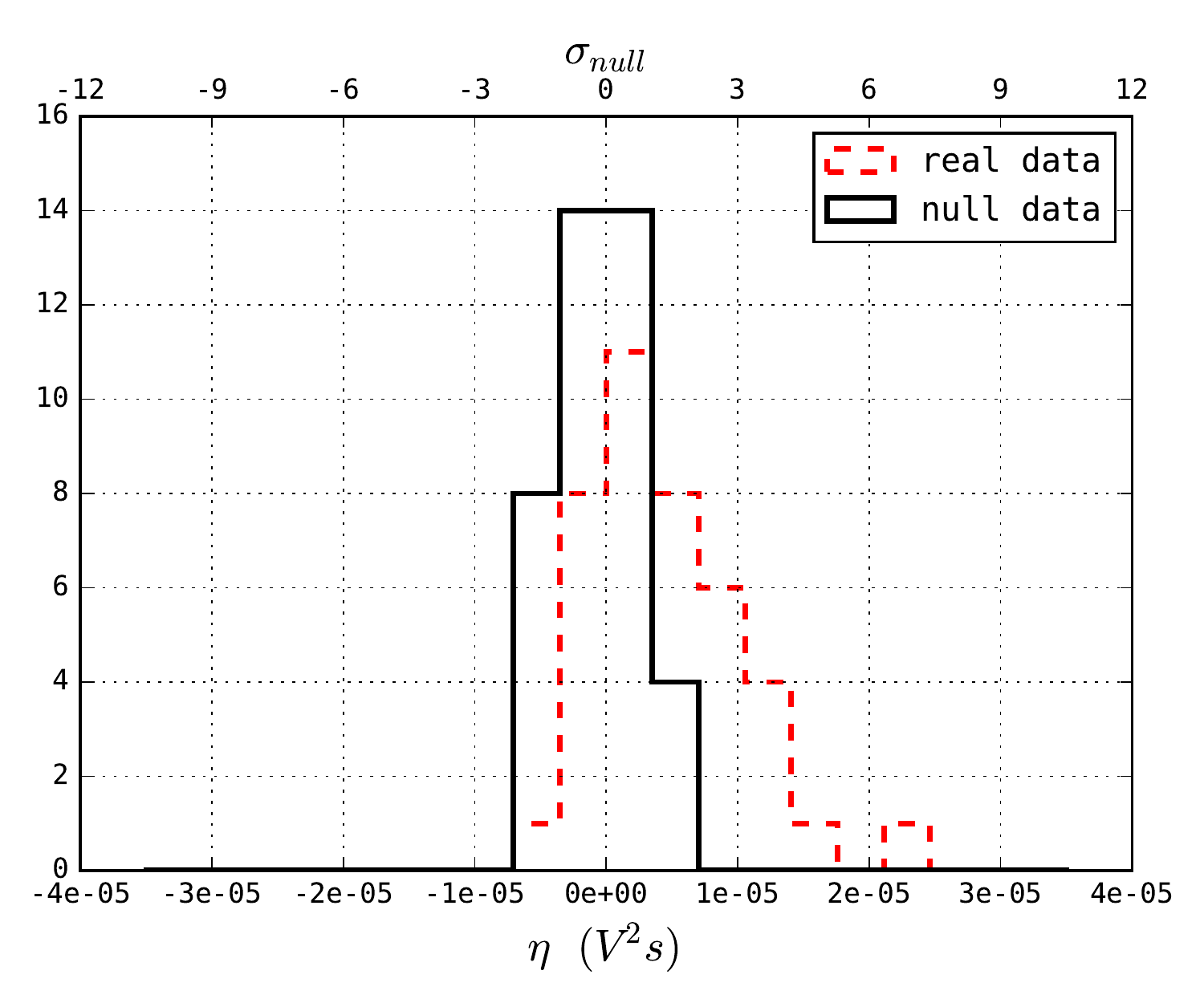}
\par\end{centering}
\caption{The significance of the excess in the signal region of Figure~\ref{expanded_2}. The significance of the mean of the real distribution is estimated to be 1.46$\sigma_{null}$.}
\label{appendix:significance_2}
\end{figure}

The results of this procedure are shown in Figure~\ref{expanded_2}. We see that the null set indeed appears to be dominated by noise, and the real set is very quiet except for a small excess in the signal region. This is again explained by the lack of phase stability in the signal region. We note that this signal excess is smaller than the excess after filtration using the null basis (Figure~\ref{expanded}). The significance of this excess is shown in Figure~\ref{appendix:significance_2}, and is significantly smaller than the same procedure using the null basis, as expected. 
 
\vfill

\include{appendix:expansion}
\end{document}